\documentclass[a4paper]{report}
\usepackage[utf8]{inputenc}
\usepackage[T1]{fontenc}
\usepackage{RJournal}
\usepackage{mathrsfs}
\usepackage{amsmath, amssymb, amsfonts, amsthm,rotating,booktabs,array}
\usepackage{bm}
\usepackage{booktabs}
\usepackage{subfigure}
\usepackage{graphicx}
\usepackage{mathabx}
\usepackage{multirow}
\usepackage{setspace}
\usepackage{algorithm}
\usepackage{mathtools}
\usepackage{multicol}
\usepackage{verbatim}

\newcommand{\noop}[1]{}

\begin{document}

\sectionhead{}
\volume{XX}
\volnumber{YY}
\year{20ZZ}
\month{AAAA}

\begin{article}
\title{RobustGaSP: Robust Gaussian Stochastic Process Emulation in  {R}}

\author{Mengyang Gu, Jes{\'u}s Palomo,  and  James O. Berger}

\maketitle

\abstract{
  Gaussian stochastic process (GaSP) emulation is a powerful tool for approximating computationally intensive computer models. However, estimation of parameters in the GaSP emulator is a challenging task. No closed-form estimator is available and many numerical problems arise with standard estimates, e.g., the maximum likelihood estimator. In this package, we implement a marginal posterior mode estimator, for special priors and parameterizations. This estimation method that meets the robust parameter estimation criteria was discussed in \cite{Gu2018robustness}; mathematical reasons are provided therein to explain why robust parameter estimation can greatly improve predictive performance of the emulator. In addition, inert inputs (inputs that almost have no effect on the variability of a function) can be identified from the marginal posterior mode estimation at  no extra computational cost. The package also implements the parallel partial Gaussian stochastic process (PP GaSP) emulator  (\cite{gu2016parallel}) for the scenario where the computer model has multiple outputs on, for example, spatial-temporal coordinates.  The package can be operated in a default mode, but also allows numerous user specifications, such as the capability of specifying trend functions and noise terms. Examples are studied herein to highlight  the performance of the package in terms of out-of-sample prediction.}

\section[Introduction]{Introduction}

A GaSP emulator is a fast surrogate model used to approximate the outcomes of a computer model (\cite{sacks1989design,bayarri2007framework,paulo2012calibration,palomo2015save,gu2016parallel}). The prediction accuracy of the emulator often depends strongly on the quality of the parameter estimates in the GaSP model. Although the mean  and variance parameters in the GaSP model are relatively easy to deal with, estimation of parameters in the correlation functions is difficult (\cite{kennedy2001bayesian}). Standard methods of estimating these parameters, such as maximum likelihood estimation (MLE), often produce unstable results leading to inferior prediction. As shown in (\cite{Gu2018robustness}), the GaSP emulator is unstable when the correlation between any two different inputs are estimated to be close to one or to zero. The former case causes a near singularity when inverting the covariance matrix (this can partially be addressed by adding a small nugget  (\cite{andrianakis2012effect})), while the latter problem happens more often and has no easy fix.

There are several packages on the Comprehensive {R} Archive Network (CRAN, \url{https://CRAN.R-project.org/}) which  implement the GaSP model based on the MLE,  including \CRANpkg{DiceKriging} (\cite{roustant2012dicekriging}), \CRANpkg{GPfit} (\cite{macdonald2015gpfit}), \CRANpkg{mleGP} (\cite{Dancik2013}),
\CRANpkg{spatial} (\cite{venables2002spatial}), and \CRANpkg{fields} (\cite{Nychka2016}). In these packages, bounds on the parameters in the correlation function are typically implemented to overcome the numerical problems with the MLE estimates. Predictions are, however, often quite sensitive to the choice of bound, which is essentially arbitrary, so this is not an appealing fix to the numerical problems.

In \cite{Gu2016thesis},  marginal posterior modes based on several objective priors are studied. It has been found that certain parameterizations result in more robust estimators than others, and, more importantly, that some parameterizations which are in common use should clearly be avoided.  Marginal posterior modes with the robust parameterization are  mathematically stable, as the posterior density  is shown to be zero  at the two problematic cases--when the correlation is nearly equal to one or to zero. This motivates the \CRANpkg{RobustGaSP} package; examples also indicate that the package results in more accurate in out-of-sample predictions than previous packages based on the MLE.  We use the \pkg{DiceKriging} package in these comparisons, because it is a state-of-the-art implementation of the MLE methodology

The \pkg{RobustGaSP} package (\cite{Gu2016RGaSPpackage}) for {R} builds a GaSP emulator with robust parameter estimation. It provides a default method with regard to a specific correlation function, a mean/trend function and an objective prior for the parameters. Users are allowed to specify them, for example, by using a different correlation and/or trend function, another prior distribution, or by adding a noise term with either a fixed or estimated variance. Although the main purpose of the \pkg{RobustGaSP} package is to do emulation/approximation of a complex function, this package can also be used in fitting the GaSP model for other purposes, such as nonparameteric regression, modeling spatial data and so on.  For computational purposes, most of the time consuming functions in the \pkg{RobustGaSP} package are implemented in {C++}.

We highlight several contributions of this work. First of all, to compute the derivative of the reference prior with a robust parametrization in (\cite{Gu2018robustness}) is computationally expensive, however this information is needed to find the posterior mode by the low-storage quasi-Newton optimization method (\cite{nocedal1980updating}).  We introduce a robust and computationally efficient prior, called the jointly robust prior (\cite{gu2018JRprior}), to approximate the reference prior in the tail rates of the posterior.  This has been implemented as a default setting in the \pkg{RobustGaSP} package.

Furthermore, the use of the jointly robust prior provides a natural shrinkage for sparsity and thus can be used to identify inert/noisy inputs (if there are any), implemented in the \code{findInertInputs} function in the \pkg{RobustGaSP} package. A formal approach to Bayesian model selection requires a comparison of $2^p$ models for $p$ variables, whereas in the \pkg{RobustGaSP} package, only the posterior mode of the full model has to be computed.  Eliminating mostly inert inputs in a computer model is similar to not including regression coefficients that have a weak effect, since the noise introduced in their estimation degrades prediction. However, as the inputs have a nonlinear effect to the output, variable selection in GaSP is typically much harder than the one in the linear regression. The \code{findInertInputs} function in the \pkg{RobustGaSP} package can be used, as a fast pre-experimental check, to separate the influential inputs and inert inputs in highly nonlinear computer model outputs.

The \pkg{RobustGaSP} package also provides some regular model checks in fitting the emulator, while the robustness in the predictive performance is the focus in \cite{Gu2018robustness}. More specifically, the leave-one-out cross validation, standardized residuals and Normal QQ-plot of the standardized residuals are implemented and will be introduced in this work.

Lastly, some computer models have multiple outputs. For example, each run of the TITAN2D simulator produces up to $10^9$ outputs of the pyroclastic flow heights over a spatial-temporal grid of coordinates (\cite{patra2005parallel,Bayarri09}). The computational complexity of building a separate GaSP emulator for the output at each grid is $O(kn^3)$, where $k$ is the number of grids and $n$ is the number of computer model runs. The package also implements another computationally more efficient emulator, called the parallel partial Gaussian stochastic process emulator,  which has the computational complexity being the maximum of $O(n^3)$ and $O(kn^2)$  (\cite{gu2016parallel}). When the number of outputs in each simulation is large, the computational cost of PP GaSP is much smaller than the separate emulator of each output.

The rest of the paper is organized as follows. In the next section, we briefly review the statistical methodology of the GaSP emulator and the robust posterior mode estimation.  In Section~\nameref{sec:overview}, we describe the structure of the package and highlight the main functions implemented in this package. In Section~\nameref{sec:numerical},  several numerical examples are provided to illustrate  the behavior of the package under different scenarios. In Section~\nameref{sec:Conclusion}, we present conclusions and briefly discuss potential extensions. Examples will be provided throughout the paper for illustrative purposes.

\section[Statistical framework]{The statistical framework}
\label{sec:framework}

\subsection{GaSP emulator}
\label{subsec:GaSP_emulator}
Prior to introducing specific functions and usage of the \pkg{RobustGaSP} package, we first review the statistical formulation of the GaSP emulator of the computer model with real-valued scalar outputs.  Let $\mathbf{x} \in \mathcal X$ denote a $p$-dimensional vector of inputs for the computer model, and let $y(\mathbf{x})$ denote the resulting simulator output, which is assumed to be real-valued  in this section.  The simulator $y(\mathbf{x})$ is viewed as an unknown function modeled by the stationary GaSP model, meaning that for any inputs $\{ \mathbf{x}_1,\ldots,\mathbf{x}_n\}$ from $ \mathcal{X} $, the likelihood is a multivariate normal distribution,
\begin{equation}
 (y(\mathbf{x}_1),\ldots,y(\mathbf{x}_n))^\top\mid \bm \mu, \,\sigma^2, \,{\mathbf R} \sim \mathcal{MN} ((\mu(\mathbf{x}_1),\ldots,\mu(\mathbf{x}_n))^\top, \sigma^2   {\mathbf R}  )\,,
 \label{equ:multinormal}
 \end{equation}
here $\mu(\cdot)$ is the mean function, $\sigma^2$ is the unknown variance parameter and $ {\mathbf R}$ is the correlation matrix. The mean function is typically modeled via regression,
\begin{equation}
\mu(\mathbf{x})= \mathbf h(\mathbf{x}){\bm \theta} = \sum^q_{t=1} h_t(\mathbf{x})\theta_t \,,
\label{equ:gp_mean}
\end{equation}
 where $\mathbf h(\mathbf{x})=\left(h_1(\mathbf{x}),h_2(\mathbf{x}),...,h_q(\mathbf{x})\right)$ is a vector of specified mean basis functions and $\theta_t$ is the unknown regression parameter for basis function $h_t(\cdot)$. In the default setting of the \pkg{RobustGaSP} package, a constant basis function is used, i.e., $h(\mathbf{x})=1$; alternatively, a general mean structure can be specified by the user (see Section~\nameref{sec:overview} for details).

The  $(i,j)$ element of $\mathbf R$ in (\ref{equ:multinormal}) is modeled through a correlation function $c(\mathbf x_i, \mathbf x_j)$. The product correlation function is often assumed in the emulation of  computer models (\cite{santner2003design}),
\begin{equation}
c(\mathbf x_i, \, \mathbf x_j)=\prod^p_{l=1}c_l( x_{il},  x_{jl}),
\label{equ:product_c}
 \end{equation}
where $c_l(\cdot,\,\cdot)$ is an one-dimensional correlation function for the $l^{th}$ coordinate of the input vector. Some frequently chosen correlation functions are implemented in the \pkg{RobustGaSP} package, listed in Table~\ref{tab:kernel}.  In order to use the power exponential covariance function, one needs to specify the roughness parameter $\alpha_l$, which is often set to be close to 2; e.g., $\alpha_l=1.9$ is advocated in \cite{Bayarri09}, which maintains an adequate  smoothness level yet avoids the numerical problems with $\alpha_l=2$.

The Mat{\'e}rn correlation is commonly used in modeling spatial data (\cite{stein2012interpolation}) and has recently been advocated for computer model emulation  (\cite{Gu2018robustness}); one benefit is that the roughness parameter of the Mat{\'e}rn correlation directly controls the smoothness of the process. For example,  the Mat{\'e}rn correlation with $\alpha_l=5/2$ results in sample paths of the GaSP that are twice differentiable, a smoothness level that is usually desirable. Obtaining this smoothness with the more common squared exponential correlation comes at a price, however, as, for large distances, the correlation drops quickly to zero. For the Mat{\'e}rn correlation with  $\alpha_l=5/2$, the natural logarithm of the correlation only decreases linearly with distance, a feature which is much better for emulation of computer models. Based on these reasons, the Mat{\'e}rn correlation with  $\alpha_l=5/2$ is the default correlation function in \pkg{RobustGaSP}. It is also the default correlation function in some other packages, such as \pkg{DiceKriging} (\cite{roustant2012dicekriging}).

 \begin{table}[t]
\begin{center}
\begin{tabular}{ll}

  \hline
  Mat{\'e}rn $\alpha=5/2$                 & $\left(1+\frac{\sqrt{5}d}{\gamma}+\frac{5d^2}{3\gamma^2}\right)\exp\left(-\frac{\sqrt{5}d}{\gamma}\right)$ \\
  Mat{\'e}rn $\alpha=3/2$                 & $\left(1+\frac{\sqrt{3}d}{\gamma}\right)\exp\left(-\frac{\sqrt{3}d}{\gamma}\right)$ \\
  Power exponential                  &  $\exp\left\{-\left(\frac{d}{\gamma}\right)^{\alpha}\right\}$, $0<\alpha\leq 2$ \\
    \hline
\end{tabular}
\end{center}
   \caption{Correlation functions currently implemented in \pkg{RobustGaSP}. Here $\gamma$ is the range parameter and $d$ is the distance between two points in each dimension. For simplicity, the subscript $l$ in Equation (\ref{equ:product_c}) has been dropped. }
   \label{tab:kernel}
\end{table}

 Since the simulator is expensive to run, we will at most be able to evaluate $y(\mathbf{x})$ at a set of design points.  Denote the chosen design inputs as $\mathbf x^{\mathcal D}=\{\mathbf x_1^{\mathcal D},\mathbf x_2^{\mathcal D},..., \mathbf x_n^{\mathcal D}\}$, where $\mathcal D \subset \mathcal X$. The resulting outcomes of the simulator are denoted as $\mathbf y^{\mathcal D}=(y^{\mathcal D}_1,y^{\mathcal D}_2,...,y^{\mathcal D}_n)^\top$. The design points are usually chosen to be ``space-filling", including the uniform design and lattice designs. The Latin hypercube (LH) design  is a ``space-filling" design that is widely used. It is defined in a rectangle whereby each sample is the only one in each axis-aligned hyperplane containing it. LH sampling for a 1-dimensional input space is equivalent to stratified sampling, and the variance of an estimator based on stratified sampling has less variance than the random sampling scheme (\cite{santner2003design}); for a multi-dimensional input space,  the projection of the LH samples on each dimension spreads out more evenly compared to simple stratified sampling. The LH design is also often used along with other constraints, e.g., the maximin Latin Hypercube maximizes the minimum Euclidean distance in the LH samples. It has been shown  that the GaSP emulator based on maximin LH samples has a clear advantage compared to the uniform design in terms of prediction (see, e.g.,  \cite{chen2016analysis}). For these reasons, we recommend the use of the LH design, rather than the uniform design or lattice designs.

\subsection{Robust parameter estimation}
\label{subsec:robust}

The parameters in a GaSP emulator include mean parameters, a variance parameter, and range parameters, denoted as $(\theta_1,..,\theta_q,\sigma^2,\gamma_1,...,\gamma_p)$. The objective prior implemented in the \pkg{RobustGaSP} package has the form
\begin{equation}
\pi(\bm \theta, \sigma^2,\bm \gamma)\propto \frac{\pi(\bm \gamma)}{\sigma^2},
\label{equ:prior}
\end{equation}
where $\pi(\bm \gamma)$ is an objective prior for the range parameters. After integrating out $(\bm \theta, \sigma^2)$  by the prior in (\ref{equ:prior}),  the marginal likelihood is
\begin{equation}
{ \mathcal{L}(\mathbf y^{\mathcal D}| \bm \gamma)} \propto  |{\mathbf R}|^{- \frac{1}{2}} |{{\mathbf h^\top(\mathbf x^{\mathcal D} ) }{ \mathbf R^{ - 1}}{\mathbf h(\mathbf x^{\mathcal D} )} }|^{- \frac{1}{2}}  \left( S^2\right)^{-(\frac{n - q}{2})} ,
\label{equ:mp}
\end{equation}
where $S^2=   { ({\mathbf{y}^{\mathcal D})^\top{\mathbf { Q} }
{\mathbf{y}^{\mathcal D}}}  }$ with $\mathbf { Q} = { \mathbf { R}^{ - 1}}\mathbf { P}$ and $\mathbf { P} = \mathbf I_n - {\mathbf h(\mathbf x^{\mathcal D} )} {\{{{\mathbf h^\top(\mathbf x^{\mathcal D} ) }{ \mathbf R^{ - 1}}{\mathbf h(\mathbf x^{\mathcal D} )} }\}^{ - 1}}  \mathbf h^\top(\mathbf x^{\mathcal D} ) {{\mathbf { R}}^{ - 1}}$, with $\mathbf I_n$ being the identity matrix of size $n$.

The reference prior $\pi^{R}(\cdot)$ and the jointly robust prior $\pi^{JR}(\cdot)$  for the range parameters with robust parameterizations implemented in the \pkg{RobustGaSP} package are listed in Table~\ref{tab:prior}. Although the computational complexity of the value of the reference prior is the same as the marginal likelihood, the derivatives of the reference prior are computationally hard.  The numerical derivative is thus computed in the package in finding the marginal posterior mode using the reference prior. Furthermore, the package  incorporates, by default, the jointly robust prior with the prior parameters $(C_1,\ldots,C_p,a,b)$ (whose values are given in Table~\ref{tab:prior}). The properties of the jointly robust prior are studied extensively in \cite{gu2018JRprior}. The jointly robust prior approximates the reference prior reasonably well with the default prior parameters, and has a close form derivatives. The jointly robust prior is a proper prior with a closed form of the normalizing constant and the first two moments.
 In addition, the posterior modes of the jointly robust prior can identify the inert inputs, as discussed in Section \ref{subsec:identification}.

\begin{table}[t]
\begin{center}
\begin{tabular}{ll}

  \hline
 $ \pi^R(\bm \gamma)$                 & $|{\mathbf I^*}({\bm \gamma})|^{1/2} $ \\
 $ \pi^R(\bm \xi)$                 & $|{\mathbf I^*}({\bm \xi})|^{1/2} $ with $\xi_l=\log(1/\gamma_l)$, for $l=1,...,p$ \\
   $ \pi^{JR}(\bm \beta)$                   &$(\sum^{p}_{l=1} C_l \beta_l)^{a} exp(-b\sum^{p}_{l=1} C_l\beta_l)$, with $\beta_l=1/\gamma_l$, for $l=1,...,p$ \\
    \hline
\end{tabular}
\end{center}
   \caption{Different priors for the parameters in the  correlation function implemented in \pkg{RobustGaSP}. Here $\mathbf I^*(\cdot)$ is the expected Fisher information matrix, after integrating out $(\bm \theta, \sigma^2)$. The default choice of the prior parameters in $\pi^{JR}(\bm \beta)$ is $a=0.2$, $b=n^{-1/p}(a+p)$, and $C_l$ equal to the mean of ${|x^{\mathcal D}_{il}-x^{\mathcal D}_{jl}|}$, for $1\leq i,j\leq n$, $i\neq j$. }
   \label{tab:prior}
\end{table}
The range parameters $(\gamma_1,...,\gamma_p)$ are estimated by the modes of the marginal posterior distribution
 \begin{equation}
 ({\hat \gamma}_1, \ldots {\hat \gamma}_p)=  \mathop{arg max }\limits_{ \gamma_1,\ldots, \gamma_p} ( \mathcal{L\mathcal}( \mathbf{y}^{\mathcal D}|{ \gamma_1}, \ldots, {\gamma_p}) \pi({\gamma_1}, \ldots, {\gamma_p})).
 \label{equ:est_gamma}
 \end{equation}
When another parameterization is used, parameters are first estimated by the posterior mode and then transformed back to obtain  $ ({\hat \gamma}_1, \ldots {\hat \gamma}_p)$.

Various functions implemented in the \pkg{RobustGaSP} package can be reused in other studies. \code{log\_marginal\_lik} and \code{log\_marginal\_lik\_deriv} give the natural logarithm of the marginal likelihood in (\ref{equ:mp}) and its directional derivatives with regard to $\bm \gamma$, respectively. The reference priors $\pi^R(\bm \gamma)$ and $\pi^R(\bm \xi)$ are not coded separately, but  \code{neg\_log\_marginal\_post\_ref} gives the negative values of the log marginal posterior distribution and thus one can use -\code{neg\_log\_marginal\_post\_ref} minus \code{log\_marginal\_lik} to get the log reference prior. The jointly robust prior   $ \pi^{JR}(\bm \beta)$   and its directional derivatives with regard to $\bm \beta$  are coded in \code{log\_approx\_ref\_prior} and \code{log\_approx\_ref\_prior\_deriv}, respectively. All these functions are not implemented in other packages and can be reused in other theoretical studies and applications.

\subsection{Prediction}
After obtaining $ \hat{\bm \gamma}$, the predictive distribution of the GaSP emulator (after marginalizing $(\bm \theta, \sigma^2)$ out) at a new input point $\mathbf x^*$ follows a student $t$ distribution
\begin{equation}
y({\mathbf x^{*}}) \mid \mathbf{y}^\mathcal{D},\, \hat{\bm \gamma} \sim \mathcal T ( \hat y({\mathbf x}^{*}),\hat{\sigma}^2c^{**},n - q)\,,
\label{equ:predictiongp}
\end{equation}
with $n-q$ degrees of freedom, where
\begin{eqnarray}
\label{equ:gppredmean}
\hat{y} ({\mathbf x}^{*}) &=& { \mathbf h({\mathbf x}^{*})} \hat{\bm{\theta}}+\mathbf{r}^\top(\mathbf{x}^*){{\mathbf R}}^{-1}\left(\mathbf{y}^\mathcal{D}-{{\mathbf{h}(\mathbf{x}^\mathcal{D} )}}\hat{\bm{\theta}}\right), \nonumber\\
\hat{\sigma}^2 &=&(n-q)^{-1}{\left(\mathbf{y}^\mathcal{D}-{\mathbf{h}(\mathbf{x}^\mathcal{D} )}\hat{\bm{\theta}}\right)}^{T}{{\mathbf R}}^{-1}\left({\mathbf{y}^\mathcal{D}}-{{\mathbf{h}}(\mathbf{x}^\mathcal{D} )}\hat{\bm{\theta}}\right), \nonumber \\
c^{**} &=& c({\mathbf x^{*}}, {\mathbf x^{*}})-{ \mathbf{r}^\top(\mathbf{x}^*){ {\mathbf R}}^{-1}\mathbf{r}(\mathbf{x}^*)} + \left({{\mathbf h(\mathbf x^{*})}-{\mathbf{h}^\top(\mathbf{x}^\mathcal{D} )}{\mathbf R}^{-1}\mathbf{r}(\mathbf{x}^*) }\right) ^\top \nonumber \\
&\hspace{+.3in}& \times \left(\mathbf{h}^\top(\mathbf{x}^\mathcal{D}){{\mathbf R}}^{-1}{\mathbf{h}(\mathbf{x}^\mathcal{D} )}\right)^{-1}\left({{\mathbf h(\mathbf x^{*})}}-{\mathbf{h}^\top(\mathbf{x}^\mathcal{D} )} {{\mathbf R}}^{-1}\mathbf{r}(\mathbf{x}^*) \right),
\end{eqnarray}
with $\hat{\bm{\theta}}=\left({{\mathbf{h}}^{T}(\mathbf{x}^\mathcal{D} )}{{\mathbf R}}^{-1} \ {{\mathbf{h}}(\mathbf{x}^\mathcal{D} )}\right)^{-1}{\mathbf{h}^\top(\mathbf{x}^\mathcal{D} )}{{\mathbf R}}^{-1}{\mathbf{y}^\mathcal{D}}$ being the generalized least squares estimator for $\bm \theta$ and $\mathbf{r}(\mathbf{x}^*) = (c(\mathbf{x}^*,{\mathbf{x}}^{\mathcal D}_1 ), \ldots, c(\mathbf{x}^*,{\mathbf{x}}^{\mathcal D}_n ))^\top$.

 The emulator interpolates the simulator at the design points $\mathbf {x}^{\mathcal{D}}_i$, $1\leq i\leq n$, because when $\mathbf x^*=\mathbf{x}^\mathcal{D}_i$, one has $\mathbf r^\top(\mathbf x^*) \mathbf R^{-1} = \mathbf e^\top_i$, where  $\mathbf e_i$ is the $n$ dimensional vector with the $i^{th}$ entry being $1$ and the others being $0$.  At other inputs, the emulator not only provides a prediction of the simulator (i.e., $\hat y({\mathbf x}^{*})$) but also an assessment of prediction accuracy. It also incorporates the uncertainty arising from estimating $\bm{\theta}$ and $\sigma^2$ since this was developed from a Bayesian perspective.

We now provide an example  in which the input has one dimension, ranging from $[0,10]$ (\cite{higdon2002space}). Estimation of the range parameters using the \pkg{RobustGaSP} package can be done through the following code:
\begin{example}
R> library(RobustGaSP)
R> library(lhs)
R> set.seed(1)
R> input <- 10 * maximinLHS(n=15, k=1)
R> output <- higdon.1.data(input)
R> model <- rgasp(design = input, response = output)
R> model
\end{example}
\begin{example}
Call:
rgasp(design = input, response = output)
Mean parameters:  0.03014553
Variance parameter:  0.5696874
Range parameters:  1.752277
Noise parameter:  0
\end{example}

\begin{figure}[t!]
\centering

	\includegraphics[scale=.8]{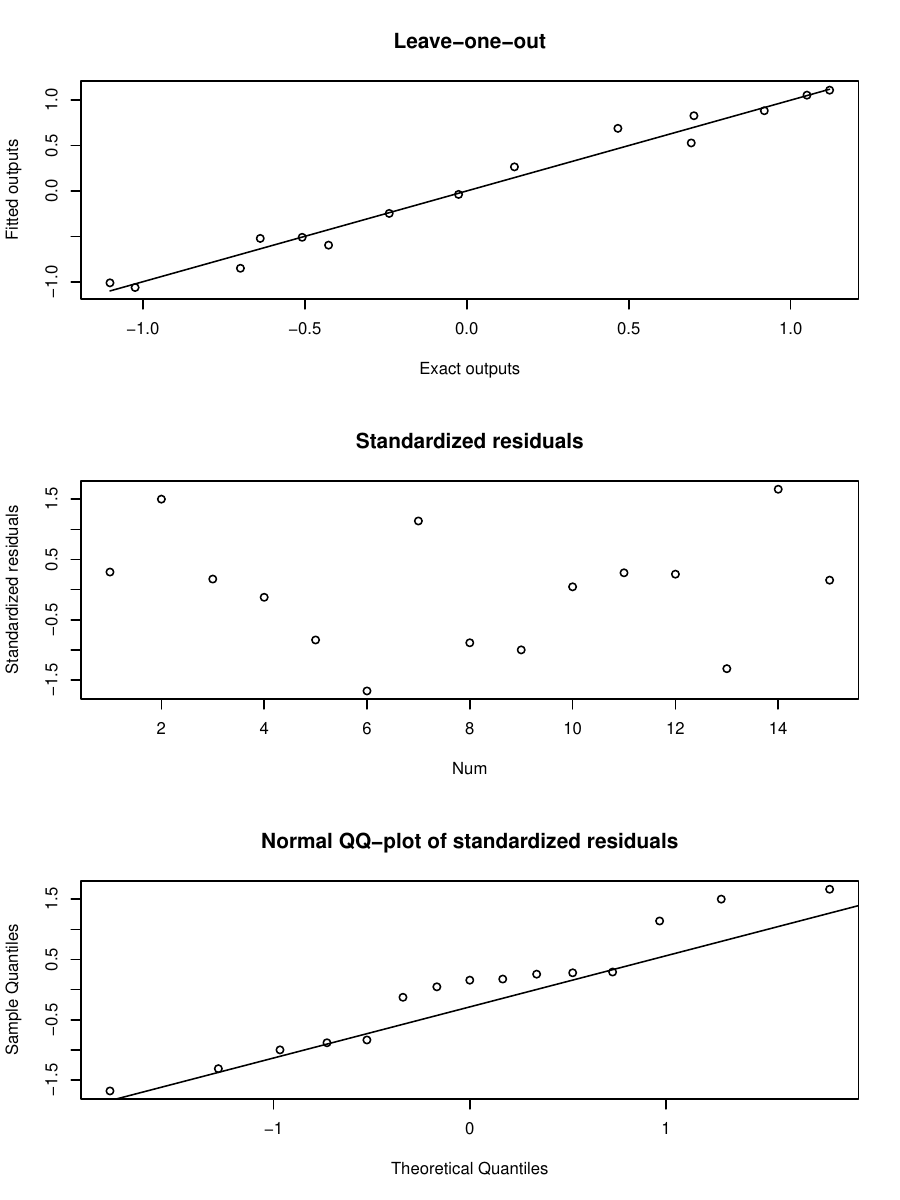}

   \caption{Leave-one-out fitted values for the GaSP model of the  \code{higdon.1.data} function in the  \pkg{RobustGaSP} package.  }

 \label{fig:Plot_eg_1}

\end{figure}

The fourth line of the code generates 15 LH samples at $[0,10]$  through the \code{maximinLHS} function of the \CRANpkg{lhs} package (\cite{lhs2016}). The function \code{higdon.1.data} is provided within the \pkg{RobustGaSP} package which has the form $y(x)=\sin(2\pi x/10)+0.2\sin(2\pi x/2.5)$. The third line fits a GaSP model with the robust parameter estimation by marginal posterior modes.

The \code{plot} function in \pkg{RobustGaSP} package implements the leave-one-out cross validation for a \code{"rgasp"} class after the GaSP model is built (see Figure~\ref{fig:Plot_eg_1} for its output):
\begin{example}
R> plot(model)
\end{example}

\begin{figure}[t]
\centering
	\includegraphics[scale=.6]{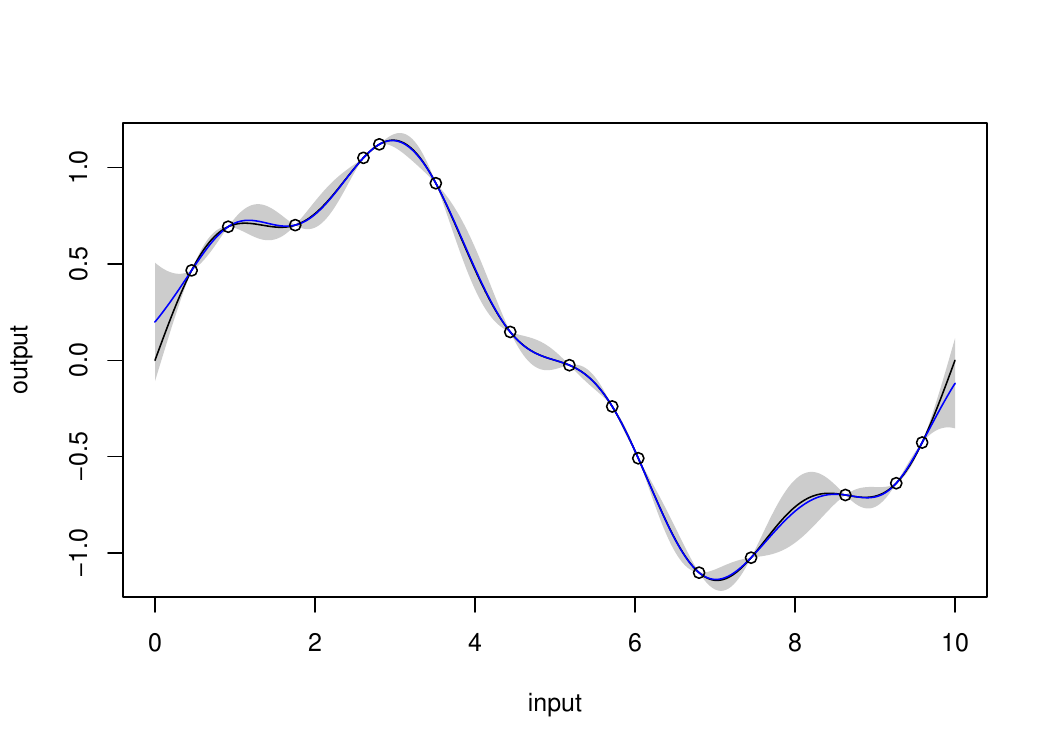}
   \caption{ The predictive mean (blue curve), the $95 \%$ predictive credible interval (grey region) and the real function (black curve). The outputs at the design points are the black circles.}
\label{figure:predict_1dim}

\end{figure}

The prediction at a set of input points can be done by the following code:
\begin{example}
R> testing_input <- as.matrix(seq(0, 10, 1/50))
R> model.predict<-predict(model, testing_input)
R> names(model.predict)
\end{example}
\begin{example}
[1] "mean"    "lower95" "upper95" "sd"
\end{example}
The \code{predict} function generates  a list containing the predictive mean, lower and upper $95\%$ quantiles and the predictive standard deviation, at each test point $\mathbf x^*$. The prediction and the real outputs are plotted in Figure~\ref{figure:predict_1dim}; produced by the following code:

\begin{example}
R> testing_output <- higdon.1.data(testing_input)
R> plot(testing_input, model.predict$mean,type='l', col='blue',
+      xlab='input', ylab='output')
R> polygon( c(testing_input,rev(testing_input)), c(model.predict$lower95,
+      rev(model.predict$upper95)), col = "grey80", border = F)
R> lines(testing_input, testing_output)
R> lines(testing_input,model.predict$mean, type='l', col='blue')
R> lines(input, output, type='p')
\end{example}

It is also possible to sample from the predictive distribution (which is a multivariate $t$ distribution) using the following code:
\begin{example}
R> model.sample <- simulate(model, testing_input, num_sample=10)
R> matplot(testing_input, model.sample, type='l', xlab='input', ylab='output')
R> lines(input, output,type='p')
\end{example}

\begin{figure}[t!]
\centering
	\includegraphics[scale=.6]{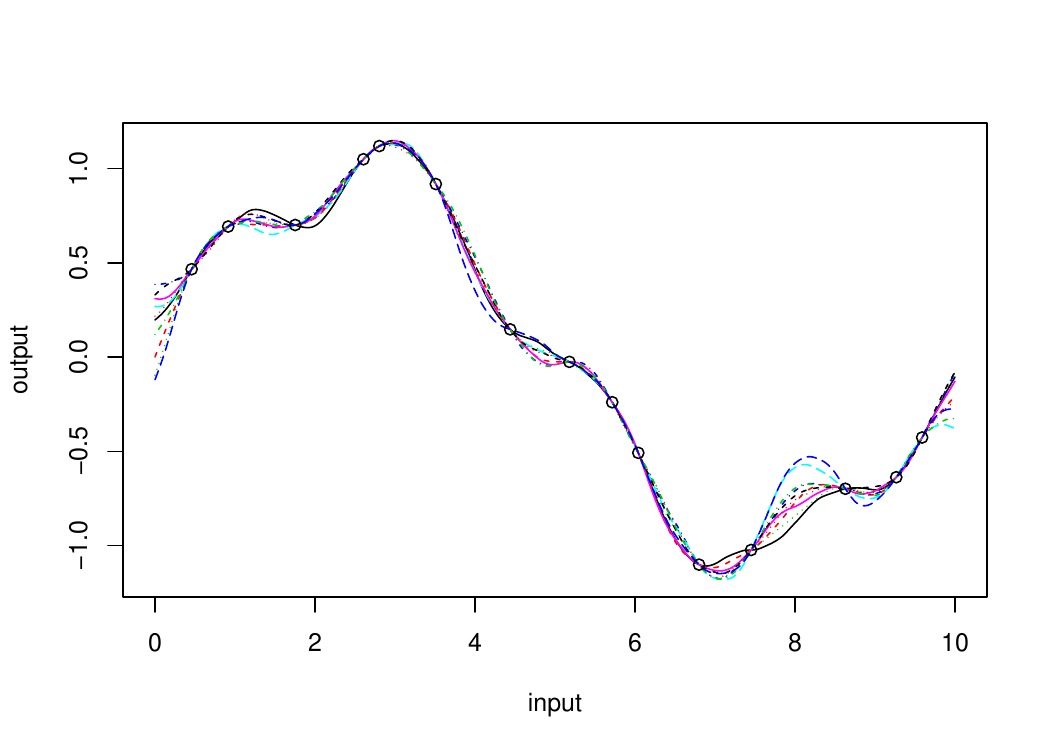}
   \caption{10 posterior predictive samples from the \pkg{RobustGaSP}. The outputs at the design points are the black circles.}
\label{figure:sample_1dim}

\end{figure}

The plots of 10 posterior predictive samples are shown in Figure~\ref{figure:sample_1dim}.

\subsection{Identification of inert inputs}
\label{subsec:identification}

Some inputs have little effect on the output of a computer model. Such inputs are called inert inputs (\cite{linkletter2006variable}).  To quantify the influence of a set of inputs on the variability of the outputs, functional analysis of the variance (functional ANOVA) can be used, often implemented through Sobol's Indices (\cite{sobol1990sensitivity,sobol2001global}).   Methods for numerical calculation of Sobol's Indices have been implemented in the \CRANpkg{sensitivity} package (\cite{sensitivity2016}) for R.

The identification of inert inputs through the posterior modes with the jointly robust prior ($\pi^{JR}(\cdot)$) for the range parameters is discussed in \cite{gu2018JRprior}. The package discussed here implements this idea, using  the {\em estimated normalized inverse range parameters},
 \begin{equation}
 \hat P_l =\frac{pC_l \hat \beta_l}{\sum^{p}_{i=1}C_i\hat \beta_i},
 \label{equ:P_l}
 \end{equation}
for $l=1,...,p$. The involvement of $C_l$ (defined in Table~\ref{tab:prior}) is to account for  the different scales of different inputs. The denominator $(\sum^{p}_{i=1}C_i\hat \beta_i)$ reflects the overall size of the estimator and $C_l \hat \beta_l$ gives the contribution of the $l^{th}$ input.  The average $\hat P_l$ is $1$ and the sum of $\hat P_l$ is $p$. When $\hat P_l$ is very close to 0, it means the $l^{th}$ input might be an inert input. In the \pkg{RobustGaSP} package, the default threshold is 0.1; i.e., when $\hat P_l<0.1$, it is suggested to be an inert input. The threshold can also be specified by users through the argument \code{threshold} in the function \code{findInertInputs}.

\begin{figure}[t]
\centering

	\includegraphics[height=.8\textwidth,width=\textwidth]{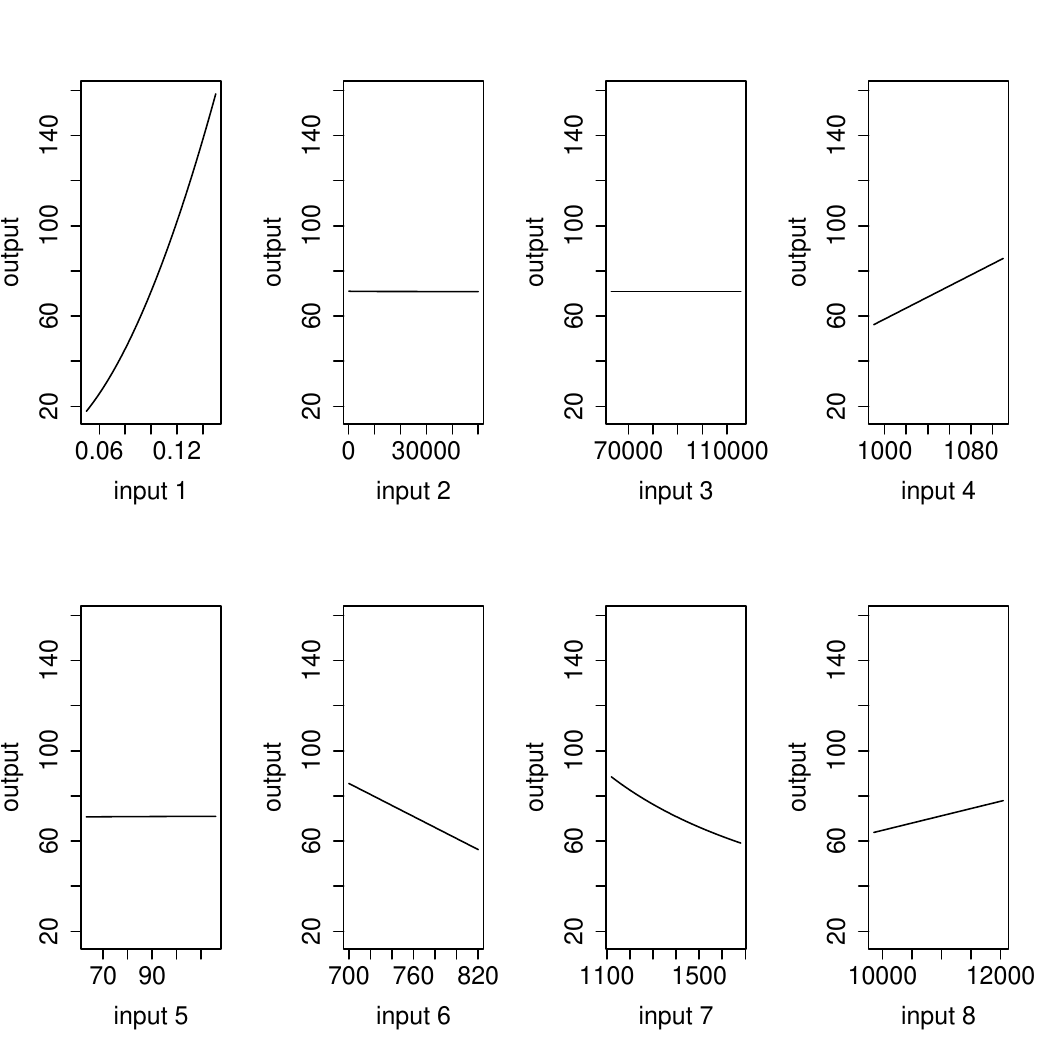}

   \caption{  Values of the borehole function by varying one input at a time. }

 \label{fig:Borehole_EDA}

\end{figure}

For demonstration purpose,  we build a GaSP emulator for the borehole experiment (\cite{worley1987deterministic,morris1993bayesian,an2001quasi}), a well-studied computer experiment benchmark which models water flow through a borehole. The output $y$ is the flow rate through the borehole in $m^3/year$ and it is determined by the equation:
 \[
 y=\frac{2\pi T_{u}(H_u-H_l)} {ln(r/r_{\omega})\big[1+\frac{2LT_u}{ln(r/r_{\omega})r^2_{\omega}K_{\omega}} +\frac{T_{u}}{T_l}\big]},
 \]
where $  r_{\omega},r, T_{u}, H_u, T_l,H_l, L$ and $K_{\omega} $ are the 8 inputs constrained in a rectangular domain with the following ranges
\begin{align*}
&r_{\omega}\in [0.05,0.15], \quad r\in[100,50000], \quad T_u\in [63070, 115600], \quad H_u \in [990, 1110],\\
&T_l \in [63.1, 116], \quad H_l \in [700, 820], \quad L \in [1120, 1680], \quad  K_{\omega} \in[9855, 12045].
\end{align*}
 We use 40 maximin LH samples to find inert inputs at the borehole function through the following code.

\begin{example}
R> set.seed(1)
R> input <- maximinLHS(n=40, k=8)  # maximin lhd sample
R> # rescale the design to the domain of the borehole function
R> LB <- c(0.05, 100, 63070, 990, 63.1, 700, 1120, 9855)
R> UB <- c(0.15, 50000, 115600, 1110, 116, 820, 1680, 12045)
R> range <- UB - LB
R> for(i in 1:8) {
R>   input[,i] = LB[i] + range[i] * input[,i]
R> }
R> num_obs <- dim(input)[1]
R> output <- matrix(0,num_obs,1)
R> for(i in 1:num_obs) {
+   output[i] <- borehole(input[i,])
+ }
R> m <- rgasp(design = input, response = output, lower_bound=FALSE)
R> P <- findInertInputs(m)
\end{example}
\begin{example}
The estimated normalized inverse range parameters are : 3.440765 8.13156e-09
4.983695e-09 0.844324 4.666519e-09 1.31081 1.903236 0.5008652
The inputs  2 3 5 are suspected to be inert inputs
\end{example}

Similar to the automatic relevance determination model in neural networks, e.g. \cite{mackay1996bayesian, neal1996bayesian}, and in machine learning, e.g. \cite{tipping2001sparse, li2002bayesian}, the function \code{findInertInputs} of the \pkg{RobustGaSP} package indicates that the $2^{nd}, 3^{rd}$, and  $5^{th}$ inputs are suspected to be inert inputs.  Figure~\ref{fig:Borehole_EDA} presents the plots of the borehole function when varying one input at a time. This analyzes the {\em local} sensitivity of an input when having the others fixed. Indeed, the output of the borehole function changes very little when the $2^{nd}, 3^{rd}$, and $5^{th}$ inputs vary.


\subsection{Noisy outputs}
\label{subsec:noise}
The ideal situation for a computer model is that it produces noise-free data, meaning that the output will not change at the same input. However, there are several cases in which the outputs are noisy. First of all, the numerical solution of the partial differential equations of a computer model could introduce small errors. Secondly, when only a subset  of inputs are analyzed, the computer model is no longer deterministic given only the subset of inputs. For example, if we only use the 5 influential inputs of the borehole function, the outcomes of this function are no longer deterministic, since the variation of the inert inputs still affects the outputs a little. Moreover, some computer models might be stochastic or have random terms in the models.


For these situations, the common adjustment is to add a  noise term to account for the error, such as $\tilde y(\cdot)=y(\cdot)+\epsilon$, where $y(\cdot)$ is the noise-free GaSP and $\epsilon$ is an i.i.d. mean-zero Gaussian white noise (\cite{ren2012objective,gu2016parallel}). To allow for marginalizing out the variance parameter,  the  covariance function for the new process $\tilde y(\cdot)$ can be parameterized as follows:
      \begin{equation}
     \sigma^2 \tilde c({{\mathbf{x}}_l},{{\mathbf{x}}_m}) {=} \sigma^2 \{c({{\mathbf{x}}_l},{{\mathbf{x}}_m}) + \eta{\delta_{lm}}\},
     \label{equ:refpriornug}
    \end{equation}
   where $\eta$ is defined to be the nugget-variance ratio and $\delta_{lm}$ is a Dirac delta function when $l=m$, $\delta_{lm}=1$. After adding the nugget, the covariance matrix becomes:
\begin{equation}
\sigma^2\tilde {\bf R}=\sigma^2({\bf R}+\eta {\bf I}_{n}).
\end{equation}
Although we call $\eta$ the nugget-variance ratio parameter, the analysis is different than when a nugget is directly added to stabilize the computation in the GaSP model. As pointed out in \cite{roustant2012dicekriging}, when a nugget is added to stabilize the computation, it is also added to the covariance function in prediction, and, hence, the resulting emulator is still an interpolator, meaning that the prediction will be exact at the design points. However, when a noise term is added, it does not go into the covariance function and the prediction at a design point will not be exact (because of the effect of the noise).

Objective  Bayesian analysis for the proposed GaSP model with the noise term can be done by defining the prior
\begin{equation}
\tilde \pi(\bm \theta, \sigma^2,\bm \gamma, \eta)\propto \frac{\tilde \pi(\bm \gamma, \eta)}{\sigma^2},
\label{equ:prior_nug}
\end{equation}
where $\tilde \pi(\bm \gamma,\eta)$ is now the prior for the range  and nugget-variance ratio parameters $(\bm \gamma, \eta)$. The reference prior and the jointly robust prior can also be extended to be $\tilde \pi^{R}(\cdot)$ and $\tilde \pi^{JR}(\cdot)$ with  robust parameterizations listed in Table~\ref{tab:prior}. Based on the computational feasibility of the derivatives and the capacity to identify noisy inputs, the proposed default setting is to use the jointly robust prior with specified prior parameters in Table~\ref{tab:prior}.

\begin{table}[t]
\begin{center}
\begin{tabular}{ll}

  \hline
 $ \tilde \pi^R(\bm \gamma, \eta)$                 & $|{\mathbf I^*}({\bm \gamma}, \eta)|^{1/2} $ \\
 $\tilde  \pi^R(\bm \xi, \eta)$                 & $|{\mathbf I^*}({\bm \xi},\eta)|^{1/2} $ with $\xi_l=\log(1/\gamma_l)$, for $l=1,...,p$ \\
   $ \tilde \pi^{JR}(\bm \beta, \eta)$                   &$(\sum^{p}_{l=1} C_l \beta_l)^{a} exp(-b(\sum^{p}_{l=1} C_l\beta_l+\eta))$, with $\beta_l=1/\gamma_l$, for $l=1,...,p$ \\
    \hline
\end{tabular}
\end{center}
   \caption{Different priors for the parameters in the correlation function implemented in \pkg{RobustGaSP}, when a noise term is present. Here $\mathbf I^*(\cdot)$ is the expected fisher information matrix after integrating out $(\bm \theta, \sigma^2)$. The default choices of the prior parameters in $\tilde \pi^{JR}(\bm \beta,\eta)$ are: $a=0.2$, $b=n^{-1/p}(a+p)$, and $C_l$ equal to the mean of ${|x^{\mathcal D}_{il}-x^{\mathcal D}_{jl}|}$, for $1\leq i,j\leq n$, $i\neq j$. }
   \label{tab:prior_nugget}
\end{table}

As in the previous noise-free GaSP model, one can estimate the range and nugget-variance ratio parameters by their marginal maximum posterior modes
 \begin{equation}
 ({\hat \gamma}_1, \ldots {\hat \gamma}_p,\hat \eta)=  \mathop{arg max }\limits_{ \gamma_1,\ldots, \gamma_p,\eta} \mathcal{L}( \mathbf{y}^{\mathcal D}|{ \gamma_1}, \ldots, {\gamma_p},\eta ) \tilde \pi({\gamma_1}, \ldots, {\gamma_p}, \eta).
 \label{equ:est_nugget_gamma}
 \end{equation}

After obtaining $\hat{ \bm \gamma}$ and $\hat \eta$, the predictive distribution of the GaSP emulator is almost the same as in Equation \eqref{equ:predictiongp}; simply replace $c(\cdot, \cdot)$ by $\tilde c(\cdot, \cdot)$ and $\mathbf R$ by $\tilde {\mathbf R}$.

Using only the influential inputs of the borehole function, we construct the GaSP emulator with a nugget based on 30 maximin LH samples through the following code:

\begin{example}
R> m.subset <- rgasp(design = input[ ,c(1,4,6,7,8)], response = output,
+    nugget.est=TRUE)
R> m.subset
\end{example}
\begin{example}

Call:
rgasp(design = input[, c(1, 4, 6, 7, 8)], response = output,
    nugget.est = TRUE)
Mean parameters:  170.9782
Variance parameter:  229820.7
Range parameters:  0.2489396 1438.028 1185.202 5880.335 44434.42
Noise parameter:  0.2265875
\end{example}

To compare  the performance of the emulator with and without a noise term, we perform some out-of-sample testing. We build the GaSP emulator by the \pkg{RobustGaSP} package and the \pkg{DiceKriging} package using  the same mean and covariance.  In  \pkg{RobustGaSP}, the parameters in the correlation functions are estimated by  marginal posterior modes with the robust parameterization, while in \pkg{DiceKriging}, parameters are  estimated by MLE with upper and lower bounds. We first construct these four emulators with the following code.

\begin{example}
R> m.full <- rgasp(design = input, response = output)
R> m.subset <- rgasp(design = input[ ,c(1,4,6,7,8)], response = output,
+    nugget.est=TRUE)
R> dk.full <- km(design = input, response = output)
R> dk.subset <- km(design = input[ ,c(1,4,6,7,8)], response = output,
+    nugget.estim=TRUE)
\end{example}

We then compare the performance of the four different emulators at 100 random inputs for the borehole function.

\begin{example}
R> set.seed(1)
R> dim_inputs <- dim(input)[2]
R> num_testing_input <- 100
R> testing_input <- matrix(runif(num_testing_input*dim_inputs),
+                          num_testing_input,dim_inputs)
R> for(i in 1:8) {
R>  testing_input[,i] <- LB[i] + range[i] * testing_input[,i]
R> }
R> m.full.predict <- predict(m.full, testing_input)
R> m.subset.predict <- predict(m.subset, testing_input[ ,c(1,4,6,7,8)])
R> dk.full.predict <- predict(dk.full, newdata = testing_input,type = 'UK')
R> dk.subset.predict <- predict(dk.subset,
+                    newdata = testing_input[ ,c(1,4,6,7,8)],type = 'UK')
R> testing_output <- matrix(0, num_testing_input, 1)
R> for(i in 1:num_testing_input) {
+   testing_output[i] <- borehole(testing_input[i, ])
+ }
R> m.full.error <- abs(m.full.predict$mean - testing_output)
R> m.subset.error <- abs(m.subset.predict$mean - testing_output)
R> dk.full.error <- abs(dk.full.predict$mean - testing_output)
R> dk.subset.error <- abs(dk.subset.predict$mean - testing_output)
\end{example}

\begin{figure}[t!]
\centering

	\includegraphics[scale=.7]{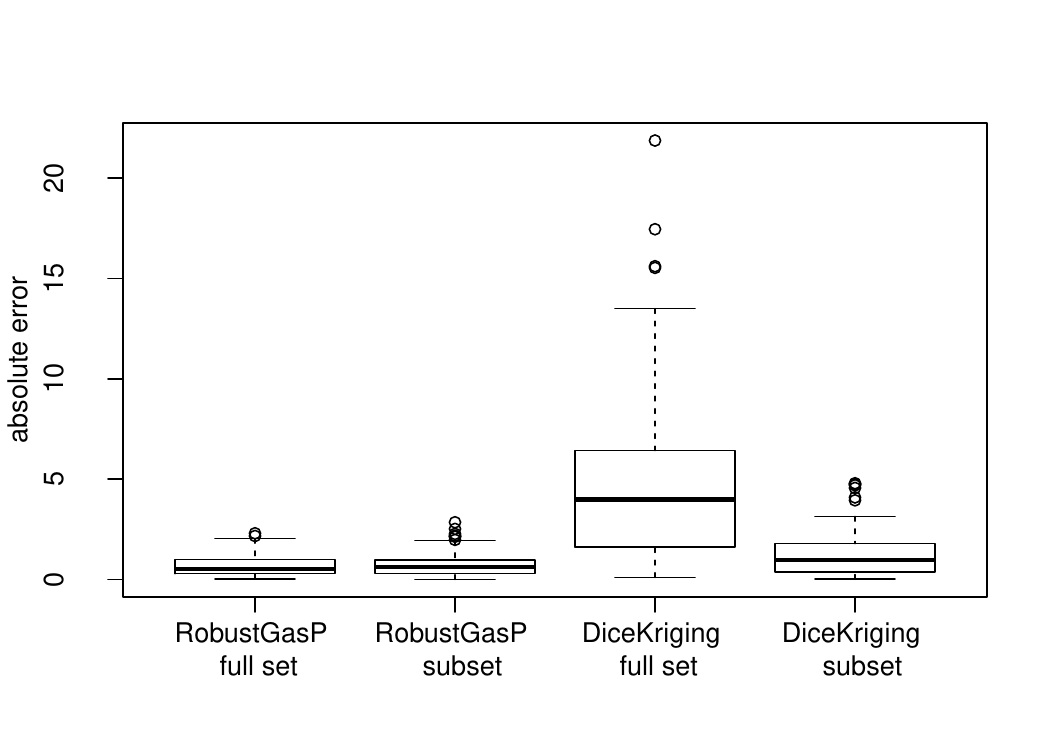}

   \caption{  Absolute out-of-sample prediction errors at 100 random samples by different emulators of the borehole function based on $n=30$ maximin LH samples. The first two boxes are the absolute predictive errors from \pkg{RobustGaSP}, with the full set of inputs and with only influential inputs (and a nugget), respectively, whereas the last two boxes  are from  \pkg{DiceKriging} with the full set of inputs and with only influential inputs (and a nugget), respectively.}

 \label{fig:Borehole_abserror}

\end{figure}


Since the \pkg{DiceKriging} package seems not to have implemented a method to estimate the noise parameter, we only compare it with the nugget case. The  box plot of the absolute errors of these 4 emulators (all with the same correlation  and mean function) at 100 held-out points  are shown in Figure~\ref{fig:Borehole_abserror}. The performance of the \pkg{RobustGaSP} package based on the full set of inputs or only influential inputs with a noise is similar, and they are both  better than the predictions from the \pkg{DiceKriging} package.



\section[An overview of {RobustGaSP} ]{An overview of \pkg{RobustGaSP}}
\label{sec:overview}
\subsection{Main functions}
\label{subsec:main_functions}
The main purpose of the \pkg{RobustGaSP} package is to predict a function at unobserved points based on only a limited number of evaluations of the function. The uncertainty associated with the predictions is obtained from the predictive distribution in Equation (\ref{equ:predictiongp}), which is implemented in two steps. The first step is to build a GaSP model through the \code{rgasp} function. This function allows users to specify the mean function, correlation function, prior distribution for the parameters, and to include a noise term or not. In the default setting, these are all specified. The mean and variance parameters are handled in a fully Bayesian way, and the range parameters in the correlation function are estimated by  their marginal posterior modes. Moreover, users can also fix the range parameters, instead of estimating them, change/replace the mean function, add a noise term, etc. The \code{rgasp} function returns an object of the \code{"rgasp"} S4 class with all needed estimated parameters, including the mean, variance, noise  and range parameters to perform predictions.

 The second step is to compute the predictive distribution of the previously created GaSP model through the \code{predict} function, which produces the  predictive means, the $95\%$  predictive credible intervals, and the predictive standard deviations at each test point. As the predictive distribution follows a student $t$ distribution in (\ref{equ:predictiongp}) for any test points, any quantile/percentile of the predictive distribution can be computed analytically. The joint distribution at a set of test points is a multivariate $t$ distribution whose dimension is equal to the number of test points. Users can also sample from the posterior predictive distribution by using the \code{simulate} function.

 The identification of inert inputs can be performed using the \code{findInertInput} function. As it only depends on the inverse range parameters through Equation (\ref{equ:P_l}), there is no extra computational cost in their identification (once the robust GaSP model has been built through the \code{rgasp} function). We suggest using the jointly robust prior by setting the argument \code{prior\_choice="ref\_approx"} in the \code{rgasp} function before calling the \code{findInertInput} function, because the penalty given by this prior is close to an $L_1$ penalty for the logarithm of the marginal likelihood (with the choice of default prior parameters) and, hence,  it can shrink the parameters for those inputs with small effect.

Besides, the  \pkg{RobustGaSP} package also implements the PP GaSP emulator introduced in \cite{gu2016parallel} for the computer model with a vector of outputs. In the PP GaSP emulator, the variances and the mean values of the computer model outputs at different grids are allowed to be different, whereas the covariance matrix of physical inputs are assumed to be the same across grids. In estimation, the variance and the mean parameters are first marginalized out with the reference priors. Then the posterior mode is used for estimating the parameters in the kernel. The \code{ppgasp} function builds a PP GaSP model, which returns an object of the \code{"ppgasp"} S4 class with all needed estimated parameters. Then the predictive distribution of PP GaSP model is computed through the \code{predict.ppgasp} function. Similar to the emulator of the output with the scalar output,  the \code{predict.ppgasp} function returns the  predictive means, the $95\%$  predictive credible intervals, and the predictive standard deviations at each test point.

\subsection[The {rgasp} function ]{The \code{rgasp} function}
\label{subsec:rgasp}

The \code{rgasp} function is one of the most important functions, as it performs the parameter estimation for the GaSP model of the computer model with a scalar output. In this section, we briefly review the implementation of the \code{rgasp} function and its optimization algorithm.

The $n\times p$ design matrix $\mathbf x^{\mathcal D}$ and the $n \times 1$ output vector $\mathbf y^{\mathcal D}$ are the only two required arguments (without default values) in the \code{rgasp} function. The default setting  in  the argument \code{trend}  is a constant function, i.e., $\mathbf h(\mathbf x^{\mathcal D})=\mathbf 1_n$.  One can also set \code{zero.mean=TRUE} in the \code{rgasp} function to assume the mean function in GaSP model is zero. By default, the GaSP model is defined to be noise-free, i.e., the noise parameter is $0$. However,  a noise term can be added with estimated or fixed variance. As the noise is parameterized following the form (\ref{equ:refpriornug}), the variance is marginalized out explicitly and the nugget-variance parameter $\eta$ is left to be estimated. This can be done by specifying the argument \code{nugget.est = T} in the \code{rgasp} function; when the nugget-variance parameter $\eta$ is known, it can be specified; e.g., $\eta=0.01$  indicates the nugget-variance ratio is equal to $0.01$ in \code{rgasp} and $\eta$ will be not be estimated with such a specification.

Two classes of priors of the form (\ref{equ:prior}),  with several different robust parameterizations,  have been implemented in the \pkg{RobustGaSP} package (see Table~\ref{tab:prior_nugget} for details). The prior that will be used is controlled by the argument \code{prior\_choice} in the \code{rgasp} function. The reference prior $\pi^R(\cdot)$ with $\bm \gamma$ (the conventional parameterization of the range parameters for the correlation functions in Table~\ref{tab:kernel}) and $\bm \xi=\log(1/\bm \gamma)$ parameterization can be specified through the arguments \code{prior\_choice="ref\_gamma"} and \code{prior\_choice="ref\_xi"}, respectively. The jointly robust prior $\pi^{JR}(\cdot)$ with the $\bm \beta=1/{ \bm\gamma}$ parameterization can be specified through the argument \code{prior\_choice="ref\_approx"}; this is the default choice used in \code{rgasp}, for the reasons discussed in Section~\nameref{sec:framework}.


The  correlation functions implemented in the \pkg{RobustGaSP} package are shown in Table~\ref{tab:kernel}, with the default setting being \code{kernel\_type = "matern\_5\_2"} in the \code{rgasp} function. The power exponential correlation function requires the specification of a {vector} of roughness parameters $\bm \alpha$ through the argument \code{alpha} in the \code{rgasp} function; the default value is $\alpha_l=1.9$ for $l=1,...,p$, as suggested in \cite{Bayarri09}.



\subsection[The {ppgasp} function ]{The \code{ppgasp} function}
\label{subsec:rgasp}

The \code{ppgasp} function performs the parameter estimation of the PP GaSP emulator for the computer model with a vector of outputs. In the \code{ppgasp} function, the output $\mathbf y^{\mathcal D}$ is a $n \times k$ matrix, where each row is the $k$-dimensional computer model outputs. The rest of the input quantities of the \code{ppgasp} function and \code{rgasp} function are the same.



Thus the \code{ppgasp} function return the estimated parameters, including $k$ estimated variance parameters, and $q\times k$ mean parameters when the mean basis has $q$ dimensions.


\subsection{The optimization algorithm}
\label{subsec:optimization}
Estimation of the range parameters $\bm \gamma$ is implemented through numerical search for the marginal posterior modes in Equation (\ref{equ:est_gamma}). The low-storage quasi-Newton optimization method (\cite{nocedal1980updating,liu1989limited}) has been used in the \code{lbfgs} function in the \CRANpkg{nloptr} package (\cite{nloptr2014}) for optimization. The  closed-form marginal likelihood, prior and their derivatives are all coded in {C++}. The maximum number of iterations and  tolerance bounds are allowed to be chosen by users with the default setting as \code{max\_eval=30} and \code{xtol\_rel=1e-5}, respectively.

Although maximum marginal posterior mode estimation with the robust parameterization eliminates the problems of the correlation matrix being estimated to be either  $\mathbf I_n$ or $\mathbf 1_n \mathbf 1^\top_n$,  the correlation matrix could still be close to these singularities in some scenarios, particularly when the sample size is very large. In such cases, we also utilize an upper bound for the range parameters $\bm \gamma$ (equivalent to a lower bound for $\bm \beta=1/{\bm \gamma}$). The derivation of this bound is discussed in the Appendix. This bound is implemented in the \code{rgasp} function through the argument \code{lower\_bound=TRUE}, and this is the default setting in \pkg{RobustGaSP}. As use of the bound is a somewhat ad hoc fix for numerical problems, we encourage users to also try the analysis without the bound; this can be done by specifying \code{lower\_bound=FALSE}. If the answers are essentially unchanged, one has more confidence that the parameter estimates are satisfactory.
Furthermore, if the purpose of the analysis is to detect inert inputs, users are also suggested to use the argument \code{lower\_bound=FALSE}  in the \code{rgasp} function.

Since the marginal posterior distribution could be multi-modal, the package allows for different initial values in the optimization by setting the argument \code{multiple\_starts=TRUE} in the \code{rgasp} function. The first default initial value for  each inverse range parameter is set to be $50$ times their default lower bounds, so the starting value will not be too close to the boundary. The second initial value for each of the inverse range parameter is set to be half of the mean of the jointly robust prior. Two initial values of the nugget-variance parameter are set to be $\eta=0.0001$ and $\eta=0.0002$ respectively.


\section[Numerical examples ]{Examples}
\label{sec:numerical}
 In this section, we present further examples of the performance of the \pkg{RobustGaSP} package, and include comparison with the \pkg{DiceKriging} package in {R}. We will use the same data, trend function, and correlation function for the comparisons. The default correlation function in both packages is the Mat{\'e}rn correlation with $\alpha=5/2$ and the default trend function is a constant function. The only difference is the method of  parameter estimation, as discussed in Section~\nameref{sec:framework}, where the \pkg{DiceKriging} package implements the MLE (by default) and the penalized MLE (PMLE) methods, \cite{DiceKrigingpackage}.

 \subsection{The modified sine wave function}
 \label{subsec:1d}


 It is expected that, for a one-dimensional function, both packages will perform well with an adequate number of design points, so we start with the function called the modified sine wave discussed in \cite{Gu2016thesis}. It has the form
 \[y=3\sin(5\pi x)+\cos(7\pi x), \]
 where $x=[0,1]$. We first perform emulation based on  12  equally spaced design points on $[0,1]$.

\begin{example}
R> sinewave <- function(x) {
+   3*sin(5*pi*x)*x + cos(7*pi*x)
+ }
R> input <- as.matrix(seq(0, 1, 1/11))
R> output <- sinewave(input)
\end{example}

 \begin{figure}[t]
\centering

	\includegraphics[scale=.7]{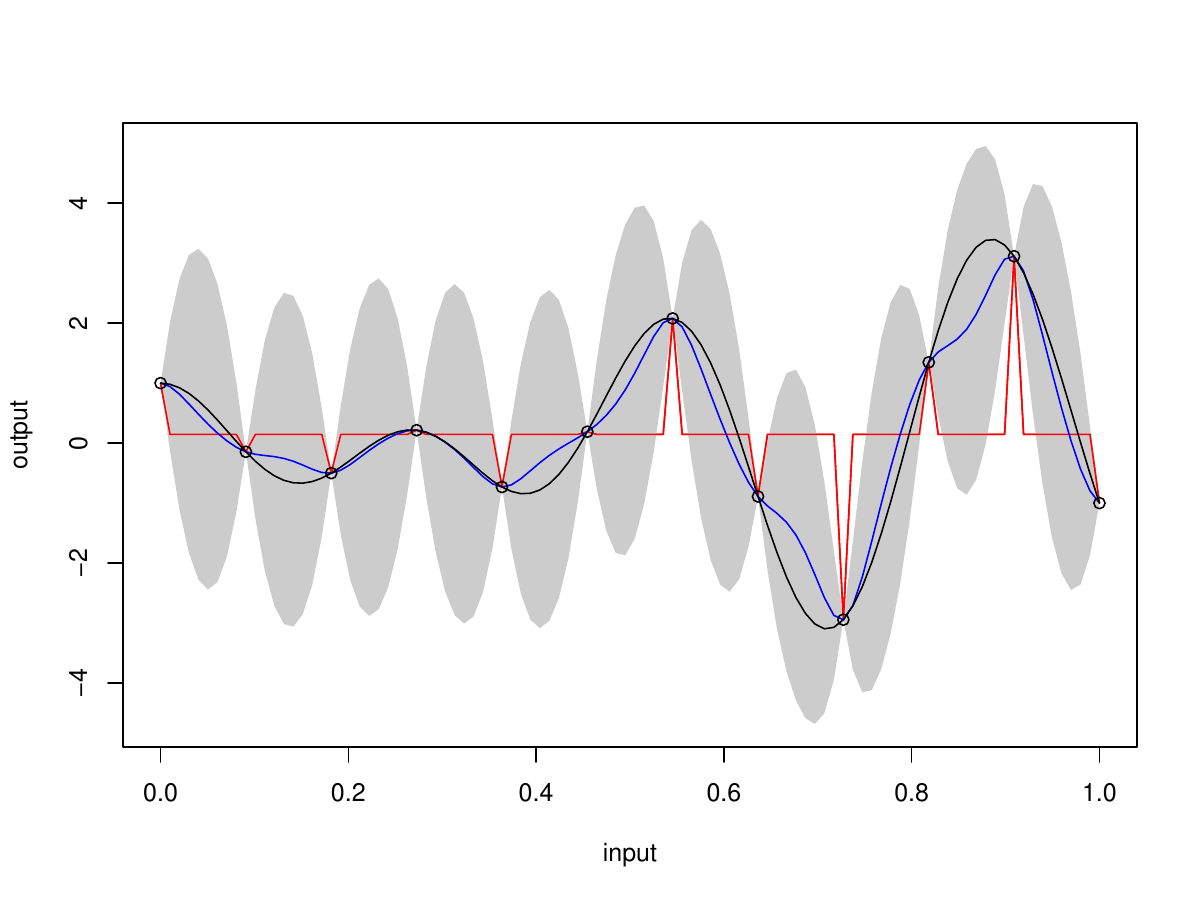}

   \caption{Emulation of the modified sine wave function with 12 design points equally spaced in $[0,1]$. The black curve is the graph of the function and the outputs at the design points are the black circles. The blue curve is the predictive mean and the grey region is the $95\%$ posterior credible interval obtained  by the \pkg{RobustGaSP} package. The red curve is the predictive mean produced by the \pkg{DiceKriging} package. }

 \label{fig:sinewave12points}

\end{figure}

The GaSP model is fitted by both the \pkg{RobustGaSP} and \pkg{DiceKriging} packages, with the constant mean function.
\begin{example}
R> m <- rgasp(design=input, response=output)
R> m
\end{example}
\begin{example}
Call:
rgasp(design = input, response = output)
Mean parameters:  0.1402334
Variance parameter:  2.603344
Range parameters:  0.04072543
Noise parameter:  0
\end{example}
\begin{example}
R> dk <- km(design = input, response = output)
R> dk
\end{example}
\begin{example}
Call:
km(design = input, response = output)

Trend  coeff.:
               Estimate
 (Intercept)     0.1443

Covar. type  : matern5_2
Covar. coeff.:
                Estimate
theta(design)     0.0000

Variance estimate: 2.327824
\end{example}
A big difference between two packages is the estimated range parameter, which is found to be around 0.04 in the \pkg{RobustGaSP} package, whereas it is found to be very close to zero in the \pkg{DiceKriging} package. To see which estimate is better, we perform prediction on 100 test points, equally spaced in $[0,1]$.
\begin{example}
R> testing_input <- as.matrix(seq(0, 1, 1/99))
R> m.predict <- predict(m, testing_input)
R> dk.predict <- predict(dk, testing_input, type='UK')
\end{example}

The emulation results are plotted in Figure~\ref{fig:sinewave12points}. Note that the red curve from the \pkg{DiceKriging} package degenerates to the fitted mean with spikes at the design points. This unsatisfying phenomenon, discussed in \cite{Gu2018robustness}, happens when the estimated covariance matrix is close to an identity matrix, i.e., $\hat {\mathbf R}\approx \mathbf I_n$, or equivalently $\hat \gamma$ tends to $0$. Repeated runs of the \pkg{DiceKriging} package under different initializations yielded essentially the same results.

The predictive mean from the \pkg{RobustGaSP} package is plotted as the blue curve in Figure~\ref{fig:sinewave12points} and is quite accurate as an estimate of the true function. Note, however, that  the uncertainty in this prediction is quite large, as shown by the wide $95\%$ posterior credible regions.

 \begin{figure}[t]
\centering

	\includegraphics[scale=.7]{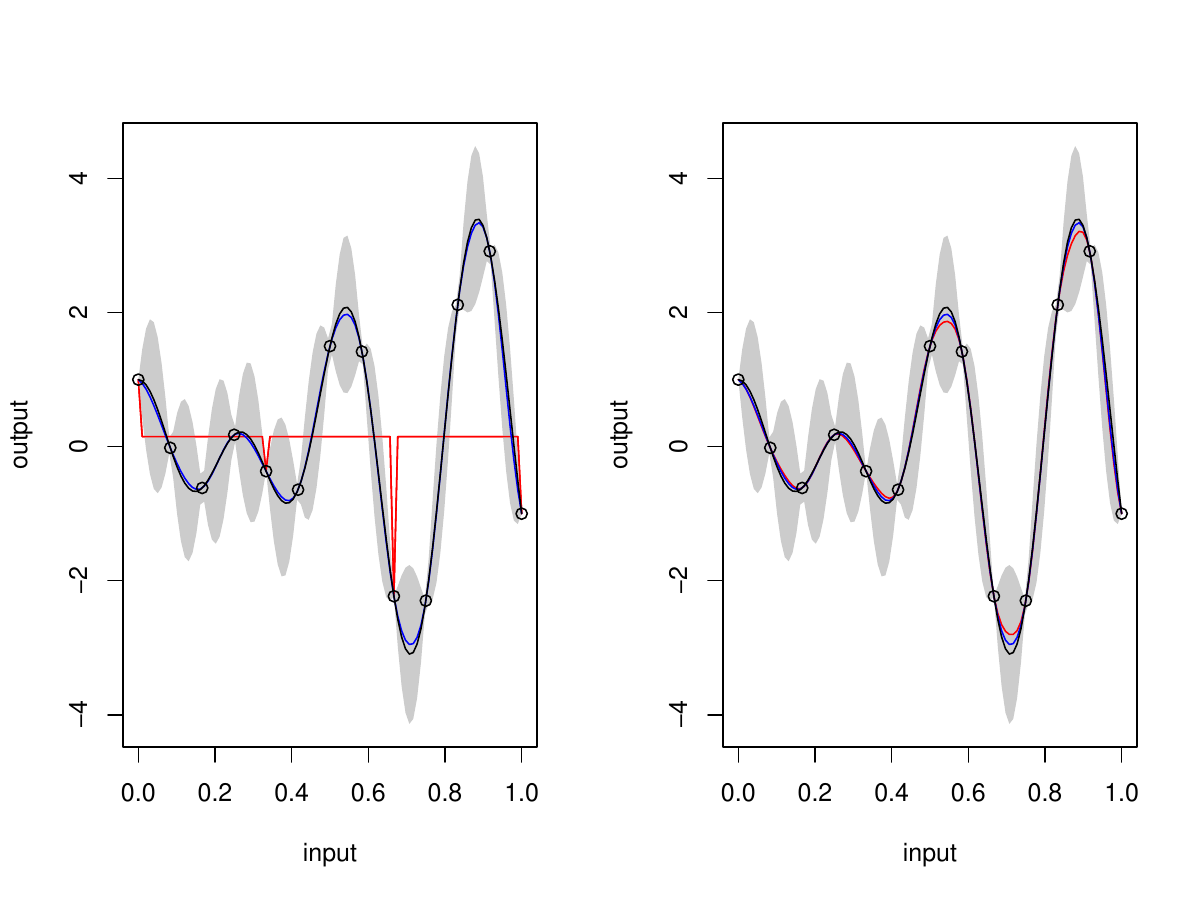}

   \caption{Emulation of the modified sine wave function with 13 design points equally spaced in $[0,1]$. The black curve is the graph of the function and the outputs at the design points are the black circles. The blue curve is the predictive mean and the grey region is the $95\%$ posterior credible interval found by \pkg{RobustGaSP}. The red curve is the predictive mean obtained by \pkg{DiceKriging}. The left panel and the right panel are two runs from \pkg{DiceKriging}, with different convergences of the optimization algorithm. }

 \label{fig:sinewave13points}
\end{figure}

In this example, adding a nugget is not helpful in \pkg{DiceKriging}, as the problem is that $\hat {\mathbf R}\approx \mathbf I_n$; adding a nugget is only helpful when the correlation estimate is close to a singular matrix (i.e., $\mathbf R\approx \mathbf 1_n \mathbf 1^\top_n$).  However,  increasing the sample size is helpful for the parameter estimation. Indeed, emulation of the modified sine wave function using 13 equally spaced design points in $[0,1]$ was successful for one run of \pkg{DiceKriging}, as shown in the right panel of Figure~\ref{fig:sinewave13points}. However, the left panel in Figure~\ref{fig:sinewave13points} gives another run of \pkg{DiceKriging} for this data, and this one converged to  the problematical $\gamma \approx 0$. The predictive mean from  \pkg{RobustGaSP} is stable. Interestingly, the uncertainty produced by \pkg{RobustGaSP} decreased markedly with the larger number of design points.

It is somewhat of a surprise that even emulation of a smooth one-dimensional function can be problematical. The difficulties with a multi-dimensional input space can be considerably greater, as indicated in the next example.


\subsection{The Friedman function}
The Friedman function was introduced in \cite{friedman1991multivariate} and is given by
\[y=10 \sin(\pi x_1x_2)+20(x_3-0.5)^2+10x_4+5x_5,\]
where $x_i \in [0,1]$ for $i=1,...,5$. 40 design points are drawn from maximin LH samples. A GaSP model is fitted using the \pkg{RobustGaSP} package and the \pkg{DiceKriging} package with the constant mean basis function (i.e., $h(\mathbf x)=1$).

\begin{example}
R> input <- maximinLHS(n=40, k=5)
R> num_obs <- dim(input)[1]
R> output <- rep(0, num_obs)
R> for(i in 1:num_obs) {
+   output[i] <- friedman.5.data(input[i,])
+ }
R> m <-rgasp(design=input, response=output)
R> dk <- km(design=input, response=output)
\end{example}

Prediction on 200 test points, uniformly sampled from $[0,1]^5$, is then  performed.

\begin{example}
R> dim_inputs <- dim(input)[2]
R> num_testing_input <- 200
R> testing_input <- matrix(runif(num_testing_input * dim_inputs),
+                         num_testing_input, dim_inputs)
R> m.predict <- predict(m, testing_input)
R> dk.predict <- predict(dk, testing_input, type='UK')
\end{example}

To compare the performance, we calculate the root mean square errors (RMSE) for both methods,
 \[\mbox{RMSE}=\sqrt{ \frac{\sum^{n^*}_{i=1}{(\hat y(\mathbf x_i^*)-y(\mathbf x_i^*) )^2}}{n^*} },\]
where $y(\mathbf x_i^*) $ is the $i^{th}$  held-out output and $\hat y(\mathbf x_i^*)$ is the prediction for $\mathbf x_i^*$ by the emulator, for $i=1,...,n^*$.
\begin{example}
R> testing_output <- matrix(0, num_testing_input, 1)
R> for(i in 1:num_testing_input) {
+   testing_output[i] < -friedman.5.data(testing_input[i,])
+ }
R> m.rmse <- sqrt(mean((m.predict$mean - testing_output)^2))
R> m.rmse
\end{example}
\begin{example}
[1] 0.2812935
\end{example}
\begin{example}
R> dk.rmse <- sqrt(mean((dk.predict$mean - testing_output)^2))
R> dk.rmse
\end{example}
\begin{example}
[1] 0.8901442
\end{example}

\begin{figure}[t]
\centering

	\includegraphics[scale=.7]{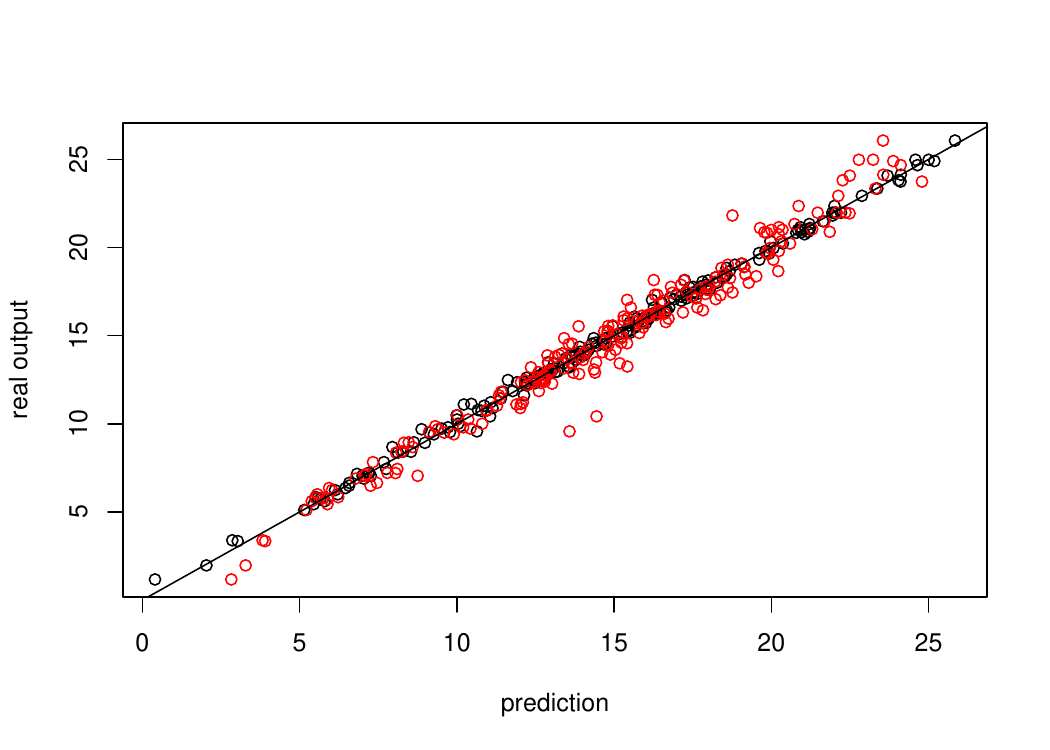}

   \caption{  Prediction of 200 held-out test points of the Friedman Function based on $40$ maximin LH samples. The y-axis is the real output and the x-axis is the prediction. The black circles are the predictive mean from  \pkg{RobustGaSP} and the red circles are the predictive mean from \pkg{DiceKriging}. A constant mean basis function is used, i.e., $h(\mathbf x)=1$. }

 \label{fig:friedman_40points}

\end{figure}

The RMSE from \pkg{RobustGaSP} is 0.28, while the RMSE from \pkg{DiceKriging} is 0.89. The predictions versus the real outputs are plotted in Figure~\ref{fig:friedman_40points}. The black circles correspond to  the predictive means from the  \pkg{RobustGaSP} package and are closer to the real output than the red circles produced by the \pkg{DiceKriging} package.
Since both packages use the same correlation  and mean  function, the only difference lies in the method of parameter estimation, especially  estimation of the range parameters, $\bm \gamma$. The \pkg{RobustGaSP} package seems to do better, leading to much smaller RMSE in out-of-sample prediction.

The Friedman function has a linear trend associated with the $4^{th}$ and the $5^{th}$ inputs (but not the first three) so we use this example to illustrate specifying a trend in the GaSP model. For realism (one rarely actually knows the trend for a computer model), we specify a linear trend for all variables; thus we use $\mathbf h(\mathbf x)=(1, \mathbf x)$, where $\mathbf x=(x_1,...,x_5)$ and investigate whether or not adding this linear trend to all inputs is helpful for the prediction.

\begin{example}
R> colnames(input) <- c("x1", "x2", "x3", "x4", "x5")
R> trend.rgasp <- cbind(rep(1, num_obs), input)
R> m.trend <- rgasp(design=input, response=output, trend=trend.rgasp)
R> dk.trend <- km(formula ~ x1 + x2 + x3 + x4 + x5, design=input, response=output)
R> colnames(testing_input) <- c("x1", "x2", "x3", "x4", "x5")
R> trend.test.rgasp <- cbind(rep(1, num_testing_input), testing_input)
R> m.trend.predict <- predict(m.trend, testing_input,
+    testing_trend=trend.test.rgasp)
R> dk.trend.predict <- predict(dk.trend, testing_input, type='UK')
R> m.trend.rmse <- sqrt(mean( (m.trend.predict$mean - testing_output)^2))
R> m.trend.rmse
\end{example}
\begin{example}
[1] 0.1259403
\end{example}
\begin{example}
R> dk.trend.rmse <- sqrt(mean((dk.trend.predict$mean - testing_output)^2))
R> dk.trend.rmse
\end{example}
\begin{example}
[1] 0.8468056
\end{example}

\begin{figure}[t]
\centering

	\includegraphics[scale=.7]{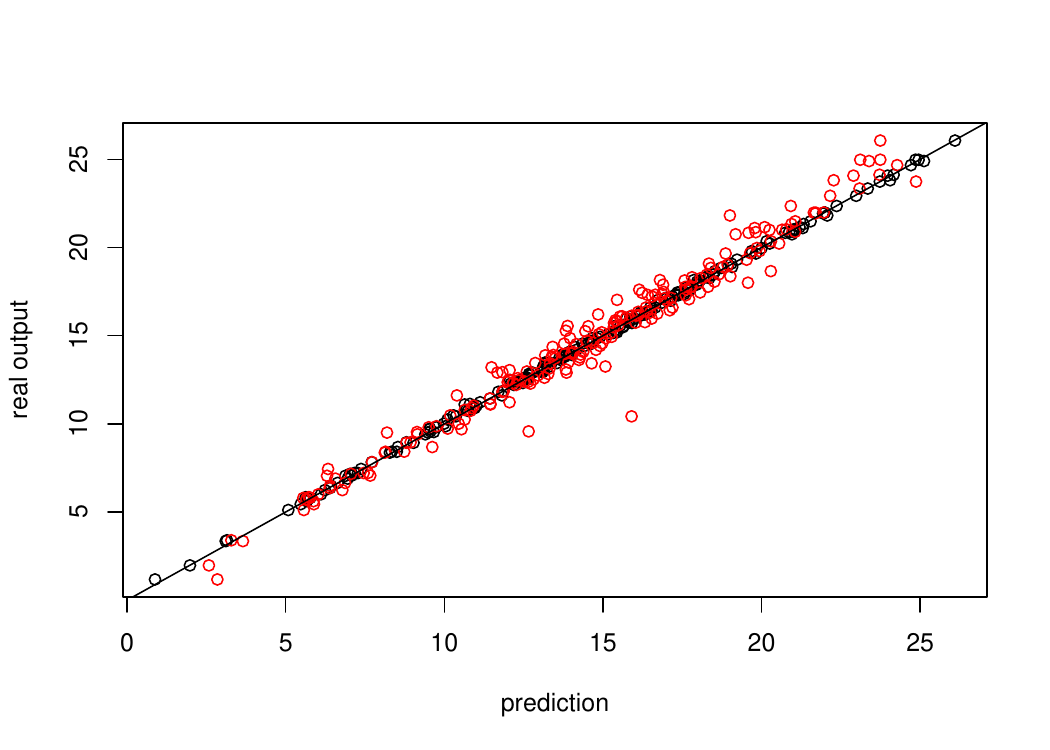}

   \caption{  Prediction of 200 held-out test points for the Friedman Function based on $40$ maximin LH design points. The y-axis is the real output and the x-axis is the prediction. The black circles are the predictive means obtained from  \pkg{RobustGaSP}, and the red circles are the predictive means obtained from the \pkg{DiceKriging} package. In both cases, linear terms are assumed for the mean basis function, i.e., $\mathbf h(\mathbf x)=(1, \mathbf x)$.}

 \label{fig:friedman_40points_trend}

\end{figure}

Adding a linear trend does improve the out-of-sample prediction accuracy of the  \pkg{RobustGaSP} package; the RMSE decreases to  $0.13$, which is only about one third of the RMSE of the previous model with the constant mean. However, the RMSE using the \pkg{DiceKriging} package with a linear mean increases to  $0.85$, more than 6 times larger than that for the \pkg{RobustGaSP}. (That the RMSE actually increased for \pkg{DiceKriging} is likely due to the additional difficulty of parameter estimation, since now the additional linear trend parameters needed to be estimated; in contrast, for \pkg{RobustGaSP}, the linear trend parameters are effectively eliminated through objective Bayesian integration.) The predictions against the real output are plotted in Figure \ref{fig:friedman_40points_trend}. The black circles correspond to the predictive means from the \pkg{RobustGaSP} package, and are an excellent match to the real outputs.

In addition to point prediction, it is of interest to evaluate the uncertainties produced by the emulators, through study of out-of-sample coverage of the resulting credible intervals and their average lengths,
\begin{align*}
{P_{CI}(95\%)} &= \frac{1}{{n^{*}}} {\sum\limits_{i = 1}^{n^{*}} 1\{y_{j}(\mathbf x^{*}_i)\in \mbox{CI}_{i}(95\% ) \}} , \\
{L_{CI}(95\%)} &= \frac{1}{n^{*}}\sum\limits_{i = 1}^{{n^{*}}} {\mbox{length}\{\mbox{CI}_{i}}(95\% )\},
\end{align*}
where $\mbox{CI}_{i}(95\% )$ is the $95\%$ posterior credible interval. An ideal emulator would have ${P_{CI}(95\%)}$ close to the $95\%$ nominal level and a short average length. We first show ${P_{CI}(95\%)} $ and ${L_{CI}(95\%)}$ for the case of a constant mean basis function.

\begin{example}
R> prop.m <- length(which((m.predict$lower95 <= testing_output)
+           & (m.predict$upper95 >= testing_output))) / num_testing_input
R> length.m <- sum(m.predict$upper95 - m.predict$lower95) / num_testing_input
R> prop.m
\end{example}
\begin{example}
[1] 0.97
\end{example}
\begin{example}
R> length.m
\end{example}
\begin{example}
[1] 1.122993
\end{example}
\begin{example}
R> prop.dk <- length(which((dk.predict$lower95 <= testing_output)
+                    & (dk.predict$upper95 >= testing_output))) / num_testing_input
R> length.dk <- sum(dk.predict$upper95 - dk.predict$lower95) / num_testing_input
R> prop.dk
\end{example}
\begin{example}
[1] 0.97
\end{example}
\begin{example}
R> length.dk
\end{example}
\begin{example}
[1] 3.176021
\end{example}

The ${P_{CI}(95\%)}$ obtained by the \pkg{RobustGaSP} is $97\%$, which is close to the $95\%$ nominal level; and $L_{CI}(95\%)$, the average lengths of the $95\%$ credible intervals, is $1.12$. In contrast, the coverage of credible intervals from \pkg{DiceKriging} is also $97\%$, but this is achieved by intervals that are, on average, about three times longer than those produced by \pkg{RobustGaSP}.

When linear terms are assumed in the basis function of the GaSP emulator, $\mathbf h(\mathbf x)=(1,\mathbf x)$,

\begin{example}
R> prop.m.trend <- length(which((m.trend.predict$lower95 <= testing_output)
+           &(m.trend.predict$upper95 >= testing_output))) / num_testing_input
R> length.m.trend <- sum(m.trend.predict$upper95 -
+                       m.trend.predict$lower95) / num_testing_input
R> prop.m.trend
\end{example}
\begin{example}
[1] 1
\end{example}
\begin{example}
R> length.m.trend
\end{example}
\begin{example}
[1] 0.8392971
\end{example}
\begin{example}
R> prop.dk.trend <- length(which((dk.trend.predict$lower95 <= testing_output)
+         & (dk.trend.predict$upper95 >= testing_output))) / num_testing_input
R> length.dk.trend <- sum(dk.trend.predict$upper95 -
+                          dk.trend.predict$lower95) / num_testing_input
R> prop.dk.trend
\end{example}
\begin{example}
[1] 0.985
\end{example}
\begin{example}
R> length.dk.trend
\end{example}
\begin{example}
[1] 3.39423
\end{example}

The ${P_{CI}(95\%)}$ for \pkg{RobustGaSP} is $100\%$ and ${L_{CI}(95\%)}=0.839$, a significant improvement over the case of a constant mean. (The coverage of $100\%$ is too high, but at least is conservative and is achieved with quite small intervals.) For \pkg{DiceKriging}, the coverage is $98.5\%$ with a linear mean, but the average interval size is now around 4 times as those produced by \pkg{RobustGaSP}.

To see whether or not the differences in performance persists when the sample size increases, the same experiment was run on the two emulators with sample size $n=80$. When the constant mean function is used, the RMSE obtained by the \pkg{RobustGaSP} package and the \pkg{DiceKriging} package were $0.05$ and $0.32$, respectively. With $\mathbf h(\mathbf x)=(1,\mathbf x)$, the RMSE's were $0.04$ and $0.34$, respectively. Thus the performance difference remains and is even larger, in a proportional sense, than when the sample size is $40$.


\subsection{DIAMOND computer model}
We illustrate the PP GaSP emulator through two computer model data sets. The first testbed is the ‘diplomatic and military operations in a non-warfighting domain’ (DIAMOND) computer model (\cite{taylor2004development}). For each given set of input variables, the dataset contains daily casualties from the 2nd and 6th day after the earthquake and volcanic eruption in Giarre and Catania. The input variables are 13-dimensional, including the speed of helicopter cruise and ground engineers, hospital and food supply capacity. The complete list of the input variables and the full data set are given in \cite{overstall2016multivariate}.

The \pkg{RobustGaSP} package includes a data set from the DIAMOND simulator, where the training and test output both contain the outputs from 120 runs of the computer model. The following code fit a PP GaSP emulator on the training data using 3 initial starting points to optimize the kernel parameters and an estimated nugget in the PP GaSP model. We then make prediction on the test inputs using the constructed PP GaSP emulator.

\begin{example}
R> data(humanity_model)
R> m.ppgasp <- ppgasp(design=humanity.X ,response=humanity.Y,
+      nugget.est=TRUE, num_initial_values=3)
R> m_pred <- predict(m.ppgasp, humanity.Xt)
R> sqrt(mean((m_pred$mean - humanity.Yt)^2))
\end{example}
\begin{example}
[1] 294.9397
\end{example}
\begin{example}
R> sd(humanity.Yt)
\end{example}
\begin{example}
[1] 10826.49
\end{example}

The predictive RMSE of the PP GaSP emulator is $294.9397$, which is much smaller than the standard deviation of the test data. Further exploration shows the output has strong positive correlation with the 11th input (food capacity). We then fit another PP GaSP model where the food capacity is included in the mean function.

\begin{example}
R> n < -dim(humanity.Y)[1]
R> n_testing=dim(humanity.Yt)[1]
R> H <- cbind(matrix(1, n, 1), humanity.X$foodC)
R> H_testing <- cbind(matrix(1, n_testing, 1), humanity.Xt$foodC)
R> m.ppgasp_trend <- ppgasp(design=humanity.X, response=humanity.Y, trend=H,
+       nugget.est=TRUE, num_initial_values=3)
R> m_pred_trend <- predict(m.ppgasp_trend, humanity.Xt, testing_trend=H_testing)
R> sqrt(mean((m_pred_trend$mean - humanity.Yt)^2))
\end{example}
\begin{example}
[1] 279.6022
\end{example}

The above result indicates the predictive RMSE of the PP GaSP emulator becomes smaller when the food capacity is included in the mean function.  We also fit GaSP emulators by the \pkg{DiceKriging} package independently for each daily output. We include the following two criteria.
\begin{eqnarray*}
{P_{CI}(95\%)} &=& \frac{1}{k{n^{*}}}\sum\limits_{i = 1}^{k} {\sum\limits_{j = 1}^{n^{*}} 1\{y^*_{i}( \mathbf x^{*}_j)\in C{I_{ij}}(95\% )\}}\,, \\
{L_{CI}(95\%)} &=& \frac{1}{{k{n^{*}}}}\sum\limits_{i = 1}^{k} \sum\limits_{j = 1}^{{n^{*}}} {\mbox{length}\{C{I_{ij}}(95\% )\} } \,,
\end{eqnarray*}
where  for $1\leq i\leq k$ and $1\leq j\leq n^*$,  $y^*_i(\mathbf x^*_j)$ is the held-out test output of the $i^{th}$  run  at the $j^{th}$ day; $\hat y^*_{i}( \mathbf x^{*}_j)$ is  the corresponding predicted  value; $C{I_{ij}}(95\% )$ is the $95\%$ predictive credible interval; and $\mbox{length}\{C{I_{ij}}(95\% )\}$ is the length of the $95\%$ predictive credible interval. An accurate emulator should have the ${P_{CI}(95\%)}$ close to the nominal $0.95$ and have small ${L_{CI}(95\%)} $ (the average length of the predictive credible interval).

\begin{table}[t]
\begin{center}
\begin{tabular}{lrrrr}
  \hline
                                                & RMSE &$P_{CI}(95\%)$ & $L_{CI}(95\%)$    \\
  \hline
  Independent GaSP emulator constant mean           &{720.16}&{0.99000}&{3678.5}  \\
  Independent GaSP emulator selected trend           &{471.10}&{0.96667}&{2189.8}  \\
  PP GaSP constant mean           &{294.94}&{0.95167}&{1138.3}  \\
  PP GaSP selected trend           &{279.60}&{0.95333}&{1120.6}  \\
  \hline

\end{tabular}
\end{center}
   \caption{Predictive performance between the independent GaSP emulator by the \pkg{DiceKriging} package (first two rows) and PP GaSP emulator by the \pkg{RobustGaSP} package (last two rows). The selected trend means the food capacity input is included in the mean function of the emulator, whereas the constant mean denotes the constant mean function. An estimated nugget is included in all methods. The baseline RMSE is $10817.47$ using the mean of the output to predict.}

   \label{tab:prediction_humanity}
\end{table}

The predictive accuracy by the independent GaSP emulator by the \pkg{DiceKriging} and the PP GaSP emulator  for the DIAMOND computer model is recorded in Table \ref{tab:prediction_humanity}. First, we noticed the predictive accuracy of both emulators seems to improve with the food capacity included in the mean function. Second, the PP GaSP seems to have much lower RMSE than the Independent GaSP emulator by the \pkg{DiceKriging} in this example, even though the kernel used in both packages are the same. One possible reason is that estimated kernel parameters by the marginal posterior mode from the \pkg{RobustGaSP} are better. Nonetheless, the PP GaSP is a restricted model, as the covariance matrix is assumed to be the same across each output variable (i.e. casualties at each day in this example). This assumption may be unsatisfying for some applications, but the improved speed in computation can be helpful. We illustrate this point by the following example for the TITAN2D computer model.


\subsection{TITAN2D computer model}
In this section, we discuss an application of emulating the massive number of outputs on the spatio-temporal grids from the TITAN2D computer model (\cite{patra2005parallel,Bayarri09}). The TITAN2D simulates the volcanic eruption from Soufri{\`e}re Hill Volcano on Montserrat island for a given set of input, selected to be the flow volume, initial flow direction, basal friction angle, and interval friction angle. The output concerned here are the maximum pyroclastic flow heights over time at each spatial grid. Since each run of the TITAN2D takes between 1 to 2 hours, the PP GaSP emulator was developed in \cite{gu2016parallel} to emulate the outputs from the TITAN2D. The data from the TITAN2D computer model can be found in \url{https://github.com/MengyangGu/TITAN2D}.

The following code will load the TITAN2D data in R:

\begin{example}
R> library(repmis)
R> source_data("https://github.com/MengyangGu/TITAN2D/blob/master/TITAN2D.rda?raw=True")
\end{example}
\begin{example}
[1] "input_variables"          "pyroclastic_flow_heights"
[3] "loc_index"
\end{example}
\begin{example}
> rownames(loc_index)
\end{example}
\begin{example}
[1] "crater"          "small_flow_area" "Belham_Valley"
\end{example}

The data contain three data frames. The input variables are a $683 \times 4$ matrix, where each row is a set of input variables for each simulated run. The output pyroclastic flow heights is a $683 \times 23040$ output matrix, where each row is the simulated maximum flow heights on $144\times 160$ grids. The index of the location has three rows, which records the index set for the crater area, small flow area and Belham Valley.

We implement the PP GaSP emulator in the \pkg{RobustGaSP} package and test on the TITAN2D data herein. We use the first $50$ runs to construct the emulator and test it on the latter $633$ runs. As argued in \cite{gu2016parallel}, almost no one is interested in the  hazard assessment in the crater area. Thus we test our emulator for two regions with habitat before. The first one is the Belham Valley (a northwest region to the crater of the  Soufri{\`e}re Hill Volcano. The second region is the ``non-crater" area, where we consider all the area after deleting the crater area. We also delete all locations where all the outputs are zero (meaning no flow hits the locations in the training data). For those locations, one may predict the flow height to be zero.

The following code will fit the PP GaSP emulator and make predictions on the Balham Valley area for each set of held out output.

\begin{example}
R> input <- input_variables[1:50, ]
R> testing_input <- input_variables[51:683, ]
R> output <- pyroclastic_flow_heights[1:50, which(loc_index[3,]==1)]
R> testing_output <- pyroclastic_flow_heights[51:683, which(loc_index[3,]==1)]
R> n=dim(output)[1]
R> n_testing <- dim(testing_output)[1]
##delete those location where all output are zero
R> index_all_zero <- NULL
R> for(i_loc in 1: dim(output)[2]) {
+   if(sum(output[ ,i_loc]==0)==50) {
+     index_all_zero <- c(index_all_zero, i_loc)
+   }
+ }
##transforming the output
R> output_log_1 <- log(output+1)
R> m.ppgasp <- ppgasp(design=input[,1:3], response=as.matrix(output_log_1[ ,-index_all_zero]),
+   trend=cbind(rep(1, n),input[,1]), nugget.est=TRUE,max_eval=100, num_initial_values=3)
R> pred_ppgasp=predict.ppgasp(m.ppgasp, testing_input[ ,1:3],
+   testing_trend=cbind(rep(1, n_testing), testing_input[,1]))
R> m_pred_ppgasp_mean <- exp(pred_ppgasp$mean)-1
R> m_pred_ppgasp_LB <- exp(pred_ppgasp$lower95)-1
R> m_pred_ppgasp_UB <- exp(pred_ppgasp$upper95)-1
R> sqrt(mean(((m_pred_ppgasp_mean - testing_output_nonallzero)^2)))
\end{example}
\begin{example}
[1] 0.2999377
\end{example}

In the above code, we fit the model using the transformed output and the first three inputs, as the fourth input (internal friction input) has almost no effect on the output. We also transform it back for prediction. As the fourth input is not used for emulation, we add a nugget to the model.  The flow volume is included to be in the mean function, as  the flow volume is positively correlated with the flow heights in all locations. These settings were used in \cite{gu2016parallel} for fitting the PP GaSP emulator to emulate the TITAN2D computer model. The only function we have not implemented in the current version of the \pkg{RobustGaSP} package is the ``periodic folding" technique for the initial flow angle, which is a periodic input. This method will appear in a future version of the package.

\begin{table}[t]
\begin{center}
\begin{tabular}{lrrrr}
  \hline
            Belham Valley                                    & RMSE &$P_{CI}(95\%)$ & $L_{CI}(95\%)$   & Time (s) \\
  \hline
  Independent GaSP emulator            &{0.30166}&{0.91100}&{0.52957} &294.43 \\
  PP GaSP                                         &{0.29994}&{ 0.93754}&{0.59474} & 4.4160 \\
  \hline
              Non-crater area                               & RMSE &$P_{CI}(95\%)$ & $L_{CI}(95\%)$   & Time (s) \\
\hline
  Independent GaSP emulator            &{0.33374}&{0.91407}&{0.53454} &1402.04 \\
  PP GaSP                                         &{0.32516}&{ 0.94855}&{0.60432} & 20.281 \\
  \hline

\end{tabular}
\end{center}
   \caption{Predictive performance between the independent GaSP emulator by the \pkg{DiceKriging} package  and PP GaSP emulator by the \pkg{RobustGaSP} package for the outputs  of the TITAN2D computer model in the Belham Valley and non-crater area. 50 runs were used to fit the emulators and the $633$ runs were used as the held-out test outputs. The   $RMSE$, $P_{CI}(95\%)$, $L_{CI}(95\%)$ and the computational time in seconds are shown in the second column to the fifth column for each method, respectively. }

   \label{tab:prediction_TITAN2D}
\end{table}

We compare the PP GaSP emulator with the independent GaSP emulator by the \pkg{DiceKriging} package with the same choice of the kernel function, mean function and transformation in the output. The PP GaSP emulator performs slightly better in terms of the predictive RMSE and the data covered in the $95\%$ predictive credible interval by the PP GaSP is also slightly closer to the nominal $95\%$ level.

The biggest difference is the computational time for these examples. The computational complexity by the independent GaSP emulator by the \pkg{DiceKriging} package is $O(kn^3)$, as it fits $k$ emulators independently for the outputs at $k$ spatial grid. In comparison, the computational complexity by the PP GaSP is the maximum of $O(n^3)$ and $O(kn^2)$. When $k \gg n$, the computational time of the PP GaSP is dominated by $O(kn^2)$, so the computational improvement in this example is thus obvious. Note that $n$ is only $50$ here. The ratio of the computational time between the independent GaSP and PP GaSP gets even larger when $n$ increases.

We have to acknowledge that, however, the PP GaSP emulator assumes the same covariance matrix across all output vector and estimate the kernel parameters using all output data.  This assumption may not be satisfied in some applications. We do not believe that the PP GaSP emulator performs uniformly better than the independent GaSP emulator. Given the computational complexity and predictive accuracy shown in the two real examples discussed in this paper, the PP GaSP emulator can be used as a fast surrogate of a computer model with massive output.

\section[Concluding remarks]{Concluding remarks}
\label{sec:Conclusion}

Computer models are widely used in many applications in science and engineering. The Gaussian stochastic process emulator provides a fast surrogate for computationally intensive computer models. The difficulty of parameter estimation in the GaSP model is well-known, as there is no closed-form, well-behaved, estimator for the correlation parameters; and poor estimation of the correlation parameters can lead to seriously inferior predictions. The \pkg{RobustGaSP} package implements marginal posterior mode estimation of these parameters for parameterizations that satisfy the ``robustness" criteria from \cite{Gu2018robustness}. Part of the advantage of this method of estimation is that the posterior has zero density for the problematic cases in which the correlation matrix is an identity matrix or the matrix or all ones.  Some frequently used estimators, such as the MLE, do not have this property.  Several examples have been provided to illustrate the use of the \pkg{RobustGaSP} package. Results of out-of-sample prediction suggest that the estimators in \pkg{RobustGaSP}, with small to moderately large sample sizes, perform considerably better than the MLE.

Although the main purpose of the \pkg{RobustGaSP} package is to emulate computationally intensive computer models,  several functions could be useful for other purposes. For example, the \code{findInertInputs} function utilizes the posterior modes to find inert  inputs at no extra computational cost than fitting the GaSP model. A noise term can be added to the GaSP model, with fixed or estimated variance, allowing \pkg{RobustGaSP} to analyze noisy data from  either computer models or, say, spatial experiments.

While posterior modes are used for estimating the correlation parameters in the current software, it might be worthwhile to implement posterior sampling for this Bayesian model. In GaSP models, the usual computational bottleneck for such sampling is the evaluation of the likelihood, as each evaluation requires inverting the covariance matrix, which is a computation of order of $O(n^3)$, with $n$ being the number of observations. As discussed in \cite{gu2016nonseparable}, however, exact evaluation of the likelihood for the Mat{\'e}rn covariance is only $O(n)$ for the case of a one-dimensional input, using the stochastic differential equation representation of the GaSP model. If this could be generalized to multi-dimensional inputs, posterior sampling would become practically relevant.

\section*{Acknowledgements}  This research was supported by NSF grants DMS-1007773, DMS-1228317, EAR-1331353, and DMS-1407775. The research of Mengyang Gu was part of his PhD thesis at Duke University.  The authors thank the editor and the referee for their comments that substantially improved the article.

\appendix 

\section*{Appendix: Likelihood functions and posterior distributions}

\bibliography{ref}

@article{li2002bayesian,
	Author = {Li, Yi and Campbell, Colin and Tipping, Michael},
	Journal = {Bioinformatics},
	Number = {10},
	Pages = {1332--1339},
	Publisher = {Oxford University Press},
	Title = {Bayesian automatic relevance determination algorithms for classifying gene expression data},
	Volume = {18},
	Year = {2002}}

@incollection{mackay1996bayesian,
	Author = {MacKay, David J.C.},
	Booktitle = {Models of neural networks III},
	Chapter = {6},
	Editor = {Eytan Domany and J. Leo van Hemmen and Klaus Schulten},
	Pages = {211--254},
	Publisher = {Springer},
	Title = {Bayesian methods for backpropagation networks},
	Year = {1996}}

@article{tipping2001sparse,
	Author = {Tipping, Michael E},
	Journal = {Journal of machine learning research},
	Number = {Jun},
	Pages = {211--244},
	Title = {Sparse Bayesian learning and the relevance vector machine},
	Volume = {1},
	Year = {2001}}

@book{neal1996bayesian,
	Author = {Neal, Radford M},
	Publisher = {Springer},
	Series = {Lecture Notes in Statistics},
	Title = {Bayesian learning for neural networks},
	Volume = {118},
	Year = {1996}}

@article{palomo2015save,
	Author = {Palomo, Jes{\'u}s and Paulo, Rui and Garc{\'\i}a-Donato, Gonzalo and others},
	Journal = {Journal of Statistical Software},
	Number = {13},
	Pages = {1--23},
	Title = {\pkg{SAVE}: an {R} package for the Statistical Analysis of Computer Models},
	Volume = {64},
	Year = {2015}}

@book{venables2002spatial,
	Author = {Venables, W.N. and Ripley, B.D.},
	Publisher = {Springer-Verlag},
	Title = {Modern Applied Statistics with {S}},
	Year = {2002}}

@manual{Nychka2016,
	Author = {Nychka, D. and Furrer, R. and Sain, S.},
	Title = {\pkg{fields}: Tools for Spatial Data. {R} package version 8.4-1},
	Url = {https://CRAN.R-project.org/package=fields},
	Year = {2016}}

@article{nocedal1980updating,
	Author = {Nocedal, Jorge},
	Journal = {Mathematics of computation},
	Number = {151},
	Pages = {773--782},
	Title = {Updating quasi-Newton matrices with limited storage},
	Volume = {35},
	Year = {1980}}

@article{sacks1989design,
	Author = {Sacks, Jerome and Welch, William J and Mitchell, Toby J and Wynn, Henry P},
	Journal = {Statistical science},
	Number = {4},
	Pages = {409--423},
	Publisher = {Institute of Mathematical Statistics},
	Title = {Design and analysis of computer experiments},
	Volume = {4},
	Year = {1989}}

@article{liu1989limited,
	Author = {Liu, Dong C and Nocedal, Jorge},
	Journal = {Mathematical programming},
	Number = {1-3},
	Pages = {503--528},
	Publisher = {Springer-Verlag},
	Title = {On the limited memory BFGS method for large scale optimization},
	Volume = {45},
	Year = {1989}}

@article{sobol1990sensitivity,
	Author = {Sobol', Il'ya Meerovich},
	Journal = {Matematicheskoe Modelirovanie},
	Number = {1},
	Pages = {112--118},
	Publisher = {Russian Academy of Sciences, Branch of Mathematical Sciences},
	Title = {On sensitivity estimation for nonlinear mathematical models},
	Volume = {2},
	Year = {1990}}

@article{friedman1991multivariate,
	Author = {Friedman, Jerome H},
	Journal = {The Annals of Statistics},
	Number = {1},
	Pages = {1--67},
	Publisher = {JSTOR},
	Title = {Multivariate adaptive regression splines},
	Volume = {19},
	Year = {1991}}

@article{morris1993bayesian,
	Author = {Morris, Max D and Mitchell, Toby J and Ylvisaker, Donald},
	Journal = {Technometrics},
	Number = {3},
	Pages = {243--255},
	Publisher = {Taylor \& Francis Group},
	Title = {Bayesian design and analysis of computer experiments: use of derivatives in surface prediction},
	Volume = {35},
	Year = {1993}}

@article{kennedy2001bayesian,
	Author = {Kennedy, Marc C and O'Hagan, Anthony},
	Journal = {Journal of the Royal Statistical Society B},
	Number = {3},
	Pages = {425--464},
	Publisher = {Wiley Online Library},
	Title = {Bayesian calibration of computer models},
	Volume = {63},
	Year = {2001}}

@article{sobol2001global,
	Author = {Sobol, Ilya M},
	Journal = {Mathematics and computers in simulation},
	Number = {1},
	Pages = {271--280},
	Publisher = {Elsevier},
	Title = {Global sensitivity indices for nonlinear mathematical models and their Monte Carlo estimates},
	Volume = {55},
	Year = {2001}}

@article{an2001quasi,
	Author = {An, Jian and Owen, Art},
	Journal = {Journal of complexity},
	Number = {4},
	Pages = {588--607},
	Publisher = {Elsevier},
	Title = {Quasi-regression},
	Volume = {17},
	Year = {2001}}

@article{higdon2002space,
	Author = {Higdon, Dave and others},
	Journal = {Quantitative methods for current environmental issues},
	Title = {Space and space-time modeling using process convolutions},
	Volume = {37--56},
	Year = {2002}}

@book{santner2003design,
	Author = {Santner, Thomas J and Williams, Brian J and Notz, William I},
	Publisher = {Springer-Verlag},
	Title = {The design and analysis of computer experiments},
	Year = {2003}}

@article{taylor2004development,
	Author = {Taylor, B and Lane, A},
	Journal = {Journal of the Operational Research Society},
	Number = {4},
	Pages = {333--339},
	Publisher = {Springer},
	Title = {Development of a novel family of military campaign simulation models},
	Volume = {55},
	Year = {2004}}

@article{patra2005parallel,
	Author = {Patra, Abani K and Bauer, AC and Nichita, CC and Pitman, E Bruce and Sheridan, MF and Bursik, M and Rupp, B and Webber, A and Stinton, AJ and Namikawa, LM and others},
	Journal = {Journal of Volcanology and Geothermal Research},
	Number = {1},
	Pages = {1--21},
	Publisher = {Elsevier},
	Title = {Parallel adaptive numerical simulation of dry avalanches over natural terrain},
	Volume = {139},
	Year = {2005}}

@article{linkletter2006variable,
	Author = {Linkletter, Crystal and Bingham, Derek and Hengartner, Nicholas and Higdon, David and Kenny, Q Ye},
	Journal = {Technometrics},
	Number = {4},
	Pages = {478-490},
	Title = {Variable selection for Gaussian process models in computer experiments},
	Volume = {48},
	Year = {2006}}

@article{bayarri2007framework,
	Author = {Bayarri, Maria J and Berger, James O and Paulo, Rui and Sacks, Jerry and Cafeo, John A and Cavendish, James and Lin, Chin-Hsu and Tu, Jian},
	Journal = {Technometrics},
	Number = {2},
	Pages = {138-154},
	Title = {A framework for validation of computer models},
	Volume = {49},
	Year = {2007}}

@article{Bayarri09,
	Author = {M. J. Bayarri and James O. Berger and Eliza S. Calder and Keith Dalbey and Simon Lunagomez and Abani K. Patra and E. Bruce Pitman and Elaine T. Spiller and Robert L. Wolpert},
	Journal = {Technometrics},
	Pages = {402-413},
	Title = {Using statistical and computer models to quantify volcanic hazards},
	Volume = {51},
	Year = {2009}}

@article{ren2012objective,
	Author = {Ren, Cuirong and Sun, Dongchu and He, Chong},
	Journal = {Journal of Statistical Planning and Inference},
	Number = {7},
	Pages = {1933--1946},
	Publisher = {Elsevier},
	Title = {Objective Bayesian analysis for a spatial model with nugget effects},
	Volume = {142},
	Year = {2012}}

@article{andrianakis2012effect,
	Author = {Andrianakis, Ioannis and Challenor, Peter G},
	Journal = {Computational Statistics \& Data Analysis},
	Number = {12},
	Pages = {4215--4228},
	Publisher = {Elsevier},
	Title = {The effect of the nugget on Gaussian process emulators of computer models},
	Volume = {56},
	Year = {2012}}

@article{paulo2012calibration,
	Author = {Paulo, Rui and Garc{\'\i}a-Donato, Gonzalo and Palomo, Jes{\'u}s},
	Journal = {Computational Statistics and Data Analysis},
	Number = {12},
	Pages = {3959--3974},
	Publisher = {Elsevier},
	Title = {Calibration of computer models with multivariate output},
	Volume = {56},
	Year = {2012}}

@article{roustant2012dicekriging,
	Author = {Olivier Roustant and David Ginsbourger and Yves Deville},
	Doi = {10.18637/jss.v051.i01},
	Issn = {1548-7660},
	Journal = {Journal of Statistical Software},
	Number = {1},
	Pages = {1--55},
	Title = {DiceKriging, DiceOptim: Two R Packages for the Analysis of Computer Experiments by Kriging-Based Metamodeling and Optimization},
	Url = {http://www.jstatsoft.org/index.php/jss/article/view/v051i01},
	Volume = {51},
	Year = {2012}}

@book{stein2012interpolation,
	Author = {Stein, Michael L},
	Publisher = {Springer-Verlag},
	Title = {Interpolation of spatial data: some theory for kriging},
	Year = {2012}}

@manual{Dancik2013,
	Author = {Garrett M. Dancik},
	Note = {R package version 3.1.4},
	Title = {mlegp: Maximum Likelihood Estimates of Gaussian Processes},
	Url = {https://CRAN.R-project.org/package=mlegp},
	Year = {2013}}

@manual{nloptr2014,
	Author = {Jelmer Ypma},
	Note = {R package version 1.0.4},
	Title = {nloptr: {R} interface to NLopt},
	Url = {https://CRAN.R-project.org/package=nloptr},
	Year = {2014}}

@article{macdonald2015gpfit,
	Author = {MacDonald, Blake and Ranjan, Pritam and Chipman, Hugh},
	Journal = {Journal of Statistical Software},
	Number = {i12},
	Publisher = {Foundation for Open Access Statistics},
	Title = {GPfit: An R package for fitting a gaussian process model to deterministic simulator outputs},
	Volume = {64},
	Year = {2015}}

@article{Gu2018robustness,
	Author = {Gu, Mengyang and Wang, Xiaojing and Berger, James O},
	Journal = {Annals of Statistics},
	Number = {6A},
	Pages = {3038--3066},
	Title = {Robust {G}aussian stochastic process emulation},
	Volume = {46},
	Year = {2018}}

@manual{Gu2016RGaSPpackage,
	Author = {Mengyang Gu and Jesus Palomo and James O Berger},
	Note = {R package version 0.5.7},
	Title = {RobustGaSP: Robust Gaussian Stochastic Process Emulation},
	Url = {https://CRAN.R-project.org/package=RobustGaSP},
	Year = {2016}}

@manual{DiceKrigingpackage,
	Author = {Olivier Roustant and David Ginsbourger and Yves Deville},
	Note = {R package version 1.5.6},
	Title = {DiceKriging: Kriging Methods for Computer Experiments},
	Url = {https://CRAN.R-project.org/package=DiceKriging},
	Year = {2018}}

@phdthesis{Gu2016thesis,
	Author = {Gu, Mengyang},
	School = {Duke University},
	Title = {Robust Uncertainty Quantification and Scalable Computation for Computer Models with Massive Output.},
	Year = {2016}}

@article{gu2016parallel,
	Author = {Gu, Mengyang and Berger, James O},
	Journal = {The Annals of Applied Statistics},
	Number = {3},
	Pages = {1317--1347},
	Publisher = {Institute of Mathematical Statistics},
	Title = {Parallel partial {G}aussian process emulation for computer models with massive output},
	Volume = {10},
	Year = {2016}}

@manual{lhs2016,
	Author = {Rob Carnell},
	Note = {R package version 0.13},
	Title = {lhs: Latin Hypercube Samples},
	Url = {https://CRAN.R-project.org/package=lhs},
	Year = {2016}}

@article{chen2016analysis,
	Author = {Chen, Hao and Loeppky, J and Sacks, Jerome and Welch, W},
	Journal = {Statistical science},
	Number = {1},
	Pages = {40--60},
	Publisher = {Institute of Mathematical Statistics},
	Title = {Analysis Methods for Computer Experiments: How to Assess and What Counts?},
	Volume = {31},
	Year = {2016}}

@manual{sensitivity2016,
	Author = {Gilles Pujol and Bertrand Iooss and Alexandre Janon with contributions from Khalid Boumhaout and Sebastien Da Veiga and Jana Fruth and Laurent Gilquin and Joseph Guillaume and Loic {Le Gratiet} and Paul Lemaitre and Bernardo Ramos and Taieb Touati and Frank Weber},
	Note = {R package version 1.12.2},
	Title = {sensitivity: Global Sensitivity Analysis of Model Outputs},
	Url = {https://CRAN.R-project.org/package=sensitivity},
	Year = {2016}}

@article{overstall2016multivariate,
	Author = {Overstall, Antony M and Woods, David C},
	Journal = {Journal of the Royal Statistical Society: Series C (Applied Statistics)},
	Number = {4},
	Pages = {483--505},
	Publisher = {Wiley Online Library},
	Title = {Multivariate emulation of computer simulators: model selection and diagnostics with application to a humanitarian relief model},
	Volume = {65},
	Year = {2016}}

@article{worley1987deterministic,
	Author = {Worley, BA},
	Journal = {Available from National Technical Information Service},
	Title = {Deterministic uncertainty analysis, ORNL-0628},
	Volume = {5285},
	Year = {1987}}

@article{gu2016nonseparable,
	Author = {Gu, Mengyang and Xu, Yanxun},
	Journal = {arXiv preprint arXiv:1711.11501},
	Title = {Nonseparable {G}aussian Stochastic Process: A Unified View and Computational Strategy},
	Year = {2017}}

@article{gu2018JRprior,
	Author = {Gu, Mengyang},
	Journal = {Bayesian Analysis, \noop{}In Press. arXiv preprint arXiv:1804.09329},
	Title = {Jointly Robust Prior for {G}aussian Stochastic Process in Emulation, Calibration and Variable Selection},
	Year = {2018}}

\address{
  Mengyang Gu\\
  Department of Statistics and Applied Probability\\
  University of California, Santa Barbara\\
  Santa Barbara, California, USA\\
   \email{michaelguzju@gmail.com}}

\address{
  Jes\'us Palomo\\
  Department of Business Administration\\
  Rey Juan Carlos University\\
  Madrid, Spain\\
 \email{jesus.palomo@urjc.es}}

\address{
  James O. Berger\\
  Department of Statistical Science \\
  Duke University \\
  Durham, North Carolina, USA \\
  \email{berger@stat.duke.edu}}
\end{article}

\end{document}